\documentclass[useAMS,usenatbib]{mn2e}
\usepackage{graphicx}
\usepackage{url}

\title[Astrometric detection of a hotspot orbiting around Sgr~A*]{Performance of astrometric detection of a hotspot orbiting on the innermost stable circular orbit of the galactic centre black hole}
\author[F. H. Vincent et al.]{F. H. Vincent,$^{1,3}$\thanks{E-mail: frederic.vincent@obspm.fr} T. Paumard,$^{1,3}$ G. Perrin,$^{1,3}$ L. Mugnier,$^{2,3}$ F. Eisenhauer$^{4}$ 
\newauthor and S. Gillessen$^{4}$ \\ 
$^{1}$Observatoire de Paris-Meudon, 5 place Jules Janssen, 92195 Meudon, France\\
$^{2}$Office National d'Etudes et de Recherches A\'erospatiales, BP72 F-92322 Ch\^atillon Cedex, France \\
$^{3}$Groupement d'Int\'er\^et Scientifique PHASE between ONERA, Observatoire de Paris, CNRS and Universit\'e Paris Diderot \\
$^{4}$ Max-Planck-Institut f\"ur Extraterrestrische Physik, 85748 Garching, Germany}
\begin{document}

\date{Accepted 2010 November 24.  Received 2010 November 16; in original form 2010 October 11}

\pagerange{\pageref{firstpage}--\pageref{lastpage}} \pubyear{2010}

\maketitle

\label{firstpage}

\begin{abstract}
  The galactic central black hole Sgr~A* exhibits outbursts of radiation in the near infrared (so-called IR flares). One model of these events consists in a hotspot orbiting on the innermost stable circular orbit (ISCO) of the hole. These outbursts can be used as a probe of the central gravitational potential. One main scientific goal of the second generation VLTI instrument \textit{GRAVITY} is to observe these flares astrometrically. Here, the astrometric precision of \textit{GRAVITY} is investigated in imaging mode, which consists in analysing the image computed from the interferometric data. The capability of the instrument to put in light the motion of a hotspot orbiting on the ISCO of our central black hole is then discussed.
   We find that \textit{GRAVITY}'s astrometric precision for a single star in imaging mode is smaller than the Schwarzschild radius of Sgr~A*. The instrument can also demonstrate that a body orbiting on the last stable orbit of the black hole is indeed moving. It yields a typical size of the orbit, if the source is as bright as $m_{\mathrm{K}}=14$.
   These results show that \textit{GRAVITY} allows one to study the close environment of Sgr~A*. Having access to the ISCO of the central massive black hole probably allows constraining general relativity in its strong regime. Moreover, if the hotspot model is appropriate, the black hole spin can be constrained.
\end{abstract}

\begin{keywords}
Instrumentation: interferometers -- Astrometry -- Galaxy: centre -- Black hole physics
\end{keywords}

\section{Introduction: testing GR in its strong regime with \textit{GRAVITY}}
\label{intro}

\textit{GRAVITY} is a second generation VLTI instrument the main goal of which is to test general relativity (GR) at the very centre of the Galaxy \citep[e.g.][see also Table~\ref{grav}]{eisenhauer08}. Observing the motions of sources orbiting in the vicinity of the dark mass Sgr~A* with great accuracy will allow probing the gravitational potential of the central massive black hole.

\begin{table}
 \centering
 \begin{minipage}{140mm}
  \caption{Main characteristics of the \textit{GRAVITY} instrument.}
  \label{grav}
  \begin{tabular}{@{}ll@{}}
 \hline
 Maximum baseline length & 143~m\\ 
Number of telescopes & 4 (all UTs)\\
Aperture of each telescope & 8.2~m \\
Wavelengths used & 1.9 - 2.5~$\mu$m\\ 
Angular resolution &  4~mas \\
Size of the total\footnote{Containing the science target and a phase reference star} field of view& 2'' \\
Size of the science\footnote{Containing only the science target} field of view & 71~mas \\
\hline
\end{tabular}
\end{minipage}
\end{table}

The star observed closest to Sgr A* so far, S2, has been monitored for more than ten years, and its orbit is known with high precision \citep{schodel02, ghez08, gillessen09}. This allowed to constrain the nature of the central attractor and to advocate with a high level of confidence that it is a black hole. However, the 15.8-year period of S2 does not allow to test relativistic effects on the orbit within a few years of observation. For instance, the advance of the peribothron will probably not be detectable before the next pericentre passage (so after nearly 20 years of cumulated observations). A powerful test of GR therefore needs to have access to stars closer in, at least ten times closer to Sgr~A* to be able to measure orbits with periods of the order of a year \citep{rubilar01, paumard08} for which the relativistic effects will be greater and detectable over a shorter period of time. Such stars, fainter than S2 and more concentrated toward Sgr~A* than the currently known S-stars, will be more confused: the highest possible resolution is then crucial to observe their orbital traces.   \citet{will08} shows that a $10 \, \mu \mathrm{as}$ astrometric precision (the goal of \textit{GRAVITY}) is sufficient over one year to measure peribothron advance, frame dragging and the quadrupole momentum, provided stars with short enough periods and high enough eccentricities are available. Such a measurement would allow testing the `no-hair' theorem of GR. 

\textit{GRAVITY} aims at detecting these innermost sources in its field of view of 71 mas FWHM diameter. The analysis of the galactic centre stellar population \citep{genzel03b, paumard06}, extrapolated to the central 71 mas, allows one to estimate the number of stars that \textit{GRAVITY} will be able to detect. Fig.~7 of \citet{genzel03b} gives a number density of around 25 stars with $10<m_{\rm{K}}<17$ per square arcsecond with angular distance to the centre less than 35 mas (i.e. inside the field of view). There are thus $25 \times \pi \times 0.035^{2} \approx 0.1$ star in \textit{GRAVITY}'s field of view. Thus, less than one star in this range of magnitude is expected to be continuously detectable (as the density function is increasing with decreasing radius, more fainter stars will be detectable, but will demand a longer integration). However, stars from further out will fly through the central 71~mas from time to time due to their large enough eccentricities, such as the three $m_{\mathrm{K}}=15$ S-stars: S2, S14 and S17 nowadays.ÊThus, considering around 3 stars of magnitude $m_{\mathrm{K}} \approx 15$ to be in the field of view is slightly optimistic, but not unrealistic. This will be the case, provided that the period of observation is well chosen (in order for the closest S-stars to be present).

Sgr~A* also exhibits powerful flares in the X-ray, NIR and sub-mm domains \citep[e.g.][]{baganoff01, ghez04, clenet05, eckart09}. The NIR events are characterized by an overall time-scale of the order of one hour, and a quasi-periodicity of roughly 20 minutes; the luminosity of the source increases by a factor of 20 \citep[see][]{hamaus09}. The typical flux density at the maximum of the flare is of the order of 8 mJy \citep{genzel03a, eckart09} which translates to approximately $m_{\mathrm{K}}=15$. The brightest flare observed to date reached $m_{\mathrm{K}}=13.5$, the median flux of the source thus increasing by a factor of 27 \citep{doddseden10}.

It must be noticed that some authors question the fact that flares can be understood as specific events. The light curve fluctuation is then interpreted as a pure red noise \citep[see e.g.][]{do09}. Following this rationale, it is no longer justified to select the `flare' part of the light curve: the quasi-periodicity is then no longer significant. This debate of the event nature of the flares is investigated in \citet{doddseden10}.

Various models are investigated to explain these flares: adiabatic expansion of a synchrotron-emitting blob of plasma \citep{yusefz06}, heating of electrons in a jet \citep{markoff01}, Rossby wave instability in the disc \citep{tagger06}, or a clump of matter heated by magnetic reconnection orbiting on the innermost stable circular orbit (ISCO) of the black hole \citep{yuan04,paumard08, hamaus09, eckart09}. To be tested, this last model requires an astrometric precision at least of the order of the angular radius of the ISCO, which is a few times the Schwarzschild radius of the black hole, i.e. a few times $10 \, \mu \mathrm{as}$. 

\textit{GRAVITY} is a beam combiner that will interfere the light of the four VLT UTs. It is designed to cope with such a challenging measurement as the follow-up of an object orbiting on the ISCO of Sgr~A*. A crucial requirement to do this is to be able to correct properly and in real time the atmospheric turbulence. Turbulence makes the phase jitter and it is necessary to define a phase reference and to maintain the fringes centered on this reference. This will be made possible by observing an unresolved (with zero intrinsic phase) reference source at the same time as the science target. The phase of the reference being null, it can be used to define the zero of the optical path delay. A dedicated fringe tracking system is used in \textit{GRAVITY} to keep the fringe around the zero level \citep[for details, see][]{gillessen10}.
 
It appears clear that astrometric precision is a key element to the realization of the GR-related science cases of \textit{GRAVITY}. As the instrument will be phase referenced, will use the four VLT UTs and will have five spectral channels in low-resolution mode, the u-v plane coverage will be large enough and the precision on the phase sufficient to consider using \textit{GRAVITY} in imaging mode, i.e. to reconstruct images from the interferometric observables. Here, it is assumed that the scientific target is made of simple continuum sources in order to be able to combine the various spectral channels.

Determining  the astrometric precision in imaging mode for several fixed stars present in the instrument's field of view is the first aim of this article. A few fixed stars (1 to 3) are placed in a square box of size 100 mas. The interferometric data acquisition is simulated, realistic noise is added, the image is then computed from the noisy visibilities and phases. The {\sevensize CLEAN} algorithm is used to deconvolve the image. A Monte Carlo analysis based on this procedure will lead to the determination of the astrometric precision. This part of the study is an extension of an earlier work by \citet{paumard08}.

The second objective is analysing the case of a variable source present in the field. The typical light curve of an IR flare is modeled and the source is placed on a circular orbit of radius 30 $\mu$as (ISCO of a Schwarzschild black hole of the mass of Sgr~A*). It will be determined wether its movement is detectable, and to what extent.

\section{Standard two-field interferometric astrometry vs imaging mode astrometry}

\subsection{Standard astrometric mode}

\begin{figure*}
\centering
	\includegraphics[width=5cm,height=5cm]{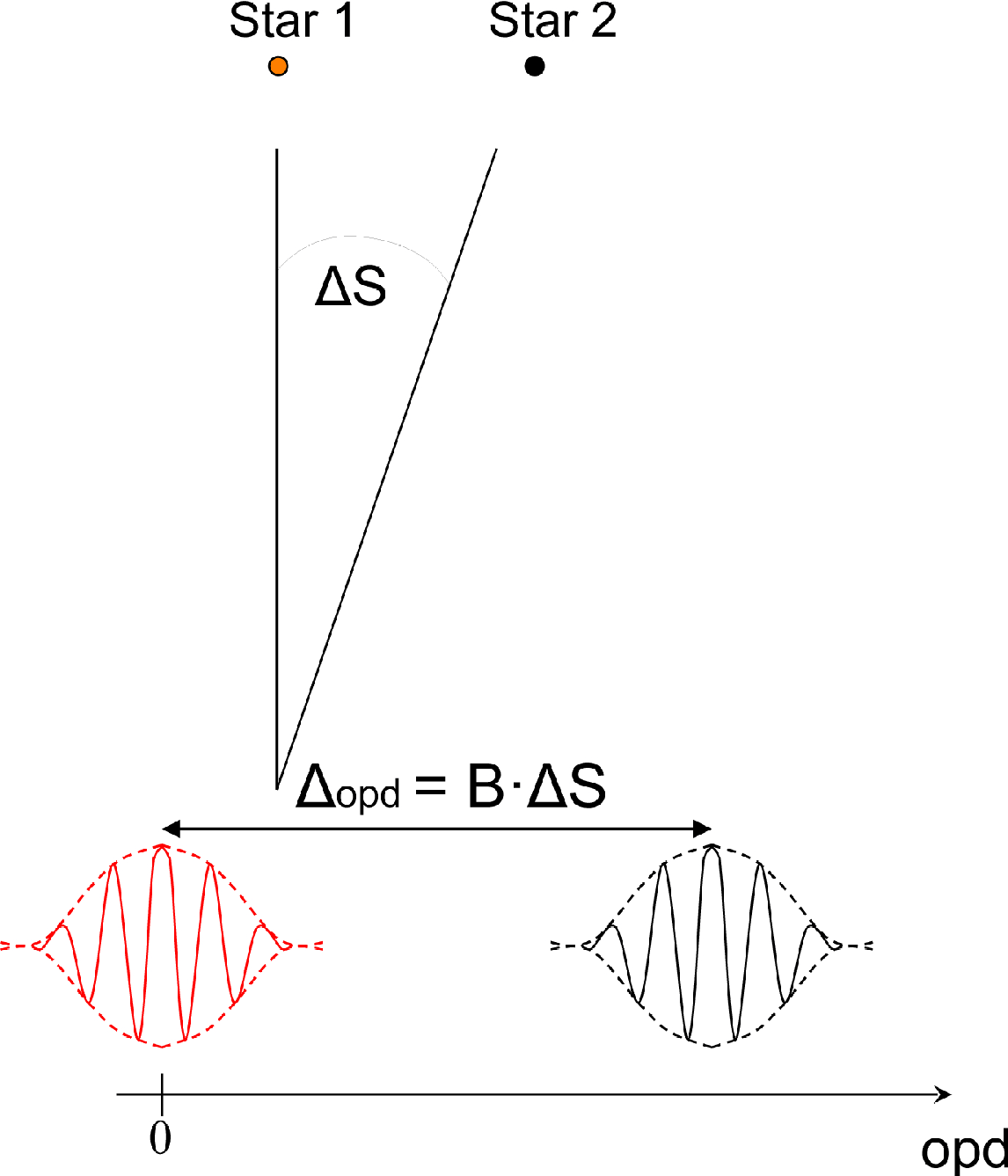}
	\caption{Standard two-field interferometric astrometry illustrated in one dimension. The baseline is $B$, the angular separation on sky is $\Delta S$, and the optical path difference (opd) is related to the phase difference through: $\Delta \phi = \frac{2 \pi}{\lambda} \Delta \rm{opd}$, where $\lambda$ is the radiation wavelength.}
	\label{stdastro}
\end{figure*}

The usual method in interferometry to determine astrometric information is to record the interferograms of a reference source and an object of interest (narrow-angle, dual-feed astrometry). If the reference and the object are supposed to be isolated (no other sources) and point sources, the phase shift between the two interferograms is simply equal to: $\Delta \phi = \Delta \phi_{\mathrm{piston}}+\frac{2\pi}{\lambda}\mathbf{B}~\mathbf{\cdot}~\Delta \mathbf{S}$, $\mathbf{B}$ being the baseline vector, $\Delta \mathbf{S}$ the angular separation on sky and $\Delta \phi_{\mathrm{piston}}$ the difference between the atmospheric piston phases affecting the two sources (which is not necessarily close to zero unless a fringe tracker is used). This is illustrated in Fig.~\ref{stdastro}, where the piston phase is not taken into account. Integrating for a time longer than the coherence time of the turbulence, the atmospheric piston will get averaged out and it will be possible to deduce the separation between the sources. However, this method is strongly dependent on the two conditions of point-like and isolated sources. If one of the sources is extended, some confusion may arise in the interpretation of the astrometric measurement as the phase information may be contaminated by the source structure, which has a non-zero intrinsic phase. Moreover, a fixed source could give rise to a changing astrometric measurement if its surface brightness distribution evolves with time. Even worse, the motion of the source components could be undetectable if the object's geometry is perfectly symmetric so that its flux barycentre is still (such as for a perfectly balanced binary), because the measured phase traces the barycentre position only. It should be noted that there is no definitive evidence (only reasonable assumptions) as to whether the environment of Sgr~A* is complex or not as this will be discovered by \textit{GRAVITY} in the near infrared with possibly first hints by \textit{PRIMA} \citep {delplancke08} and \textit{ASTRA} \citep{pott08}.

This standard method can still be used in more complex contexts by resorting to model-fitting. The complex visibility is computed assuming a given model described by some parameters (e.g. the astrometric positions of the sources and their relative intensities). Fitting these modeled visibilities to the observed ones allows one to determine the parameters. However, if the field is too complex, the number of free parameters will make it extremely difficult to use this procedure. 

\subsection{Imaging mode}

Here, the `imaging' mode of the instrument will be investigated. 

The aim is to reconstruct an image from the complex visibilities, using the Zernike--van Cittert theorem. It is thus necessary to have access to the source phase with high precision. This is made possible by using a phase reference, defining the zero of optical path delay. This procedure gives rise to specific errors on the phase, that are described in detail in Appendix~\ref{noise_Vphi} (see in particular equation (\ref{vphi_sc})).

The performance of the technique relies also on the ability to reconstruct a deconvolved image from the rough data obtained by applying the Zernike--van Cittert theorem. If the field contains only point sources, the {\sevensize CLEAN} algorithm \cite[see][]{hogbom74} will be efficient for the deconvolution\footnote{Here, a Yorick implementation of {\sevensize CLEAN} is used, which can be found at the URL \url{http://www.lesia.obspm.fr/perso/thibaut-paumard/yorick/CLEAN.i}}. The astrometric information is then directly available on the deconvolved image.

\subsection{Imaging mode or fitting?}
\label{imorfit}

The first aim of this work is to investigate the astrometric precision of such an instrument as \textit{GRAVITY} in imaging mode and in standard fitting mode. We are interesting in knowing whether the pure imaging mode can be of any scientific interest, or whether a fitting procedure is always necessary in order to obtain high enough precision.

Should the imaging mode be able to give precise enough results, it could appear as a very interesting alternative to fitting. Indeed, if the astrophysical source is too complex to allow for a few-parameter model to be determined, astrometric parameters could still be measurable directly on the image. Moreover, an image better catches most of the object's high-frequency details \citep{haubois09}. However, testing the imaging mode with a complex field of view is beyond the scope of this work as it would need a far more sophisticated algorithm than {\sevensize CLEAN}, such as {\sevensize WISARD} \citep{meimon09} or {\sevensize MIRA} \citep{thiebaut08}. A discussion of the use of these two algorithms, in comparison to standard parametric model-fitting, can be found in \citet{lebesnerais08}.

Should a fitting procedure be necessary in order to obtain the desired precision, imaging mode could still be very interesting to determine the broad lines of the field of view. This information could then be used as a first guess for the fitting.

\subsection{Limiting precision}

Whatever the method used, it is possible to compute the limiting precision that one can hope to obtain, which corresponds to the theoretical precision of the instrument. The phase error associated with an integration time of 100~s on a $m_{\rm{K}}=12$ source has been determined by simulations: $\sigma_{\phi}~=~0.0085$ rad. Then, for $N$= 30 u-v points (equal to the number of wavelenghts - 5 - multiplied by the number of baselines - 6 -), a mean baseline of $B$=100~m, and a mean wavelength of $\lambda=2.2\, \times 10^{-6}$~m, the astrometric error is of: $\sigma_{\rm{theo}} = \frac{\lambda}{2\pi \, B} \frac{\sigma_{\phi}}{\sqrt{N}} \sim 1.12$~$\mu$as. This estimate assumes that all the u-v points are aligned in one given direction. A rough estimation of the error in a given direction is: $\sigma_{\rm{y\;theo}} \approx \sqrt{2} \, \sigma_{\rm{theo}} =  1.58~\mu$as.

Thus in any case, the precision will stay above approximately 1.6 $\mu$as.

\section{Simulating the data}

\label{method}

The astrophysical object of interest, a star field, is described by a set of coordinates in mas $(x_{i}^{\mathrm{model}},y_{i}^{\mathrm{model}})$, randomly placed in a 100 mas square box, and by intensities $I_{i}$. The $(u,v)$ plane coverage is computed and the modulus and phase of visibilities are derived by means of the Zernike--Van Cittert theorem. All the different sources of noise introduced in Appendix~\ref{noise} that affect the visibility modulus and phase are then added, as a function of the source magnitude. The noisy visibilities are then transformed back to an image. This is performed by computing a numerical Fourier transform:

\begin{equation}
\mathbf{I} = \sum_{k=1}^{N_{\mathrm{uv}}} V(k)e^{i\,\varphi(k)} e^{i\left(u(k)\mathbf{x}+v(k)\mathbf{y}\right)}
\end{equation}
 where the bold quantities are 2D arrays (\textbf{I} being the image), $Ve^{i\varphi}$ is the complex visibility, and $N_{\mathrm{uv}}$ is the number of pairs of spatial frequencies (u,v).

 Because of the use of single-mode fibers, the transmission of the 71 mas FWHM field of view is gaussian: stars far from the centre will be damped. Images are deconvolved  using the {\sevensize CLEAN} algorithm, which concentrates the flux of each star in a set of a few pixels of coordinates $(x_{k},y_{k})$. The deconvolved position $(x^{\mathrm{deconv}},y^{\mathrm{deconv}})$  associated with each set of pixels is defined by computing the barycentre of the pixels weighted by the intensity. The typical number of illuminated pixels varies from 2 to 12 (depending on the number of sources and integration time) in the different cases considered in this work. Thus:

\begin{eqnarray}
	\label{bary}
x^{\mathrm{deconv}} &= & \frac{\sum\limits_{\mathrm{k}} x_{k}\,I_{k}}{I_{\mathrm{tot}}}  \\ \nonumber
y^{\mathrm{deconv}} & =& \frac{\sum\limits_{\mathrm{k}} y_{k}\,I_{k}}{I_{\mathrm{tot}}}  \nonumber
\end{eqnarray}
where $I_{k}$ is the intensity of pixel number $k$ and $I_{\mathrm{tot}}$ denotes the total intensity of the pixel set.

\begin{figure*}
\centering
	\includegraphics[width=10cm,height=10cm]{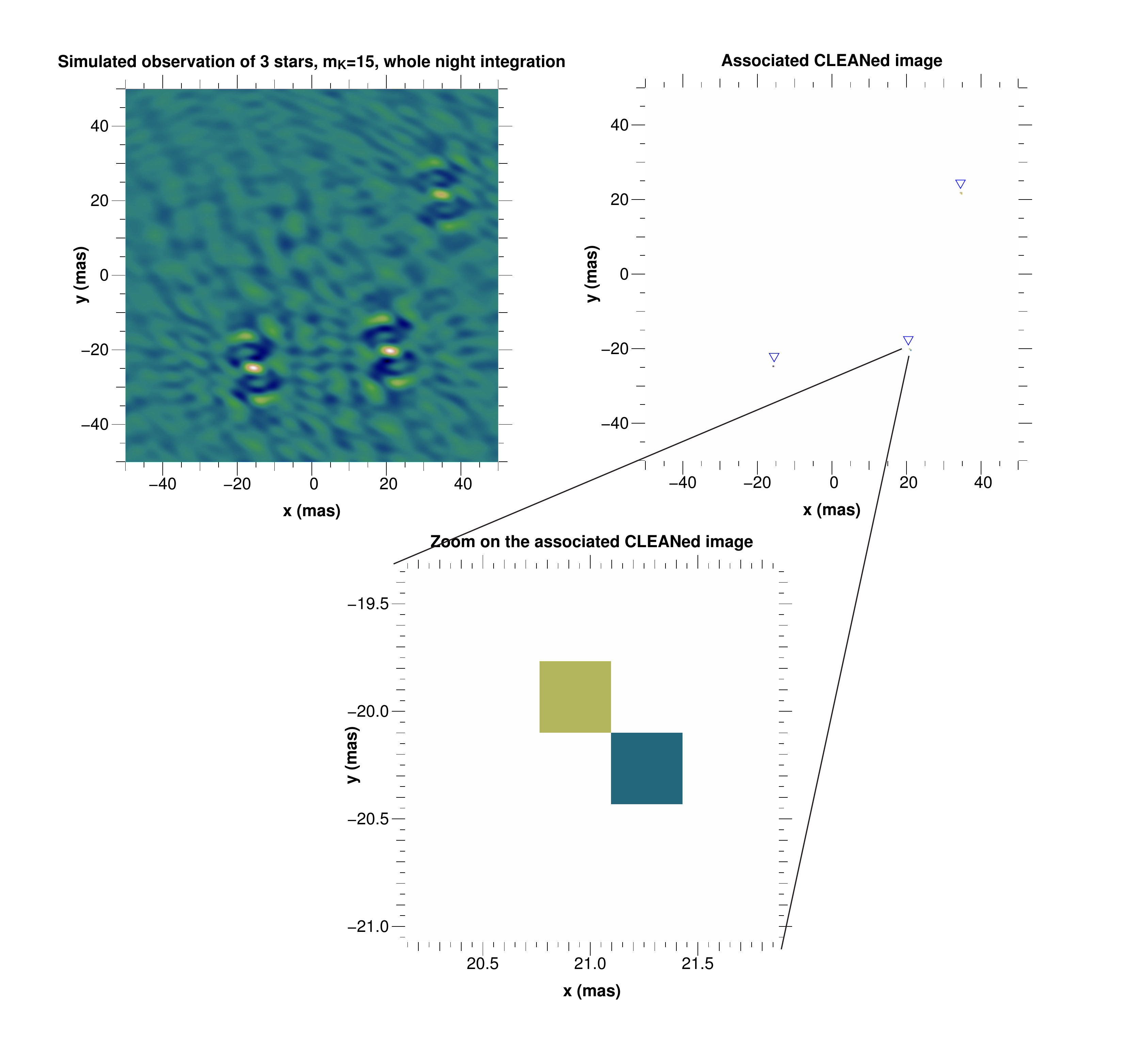}
	\caption{Top left panel: image of a simulated observation of three stars with integrated magnitude $m_{\rm{K}}=15$ made over one complete night. Top right panel: illuminated pixels obtained after deconvolution of the image by the {\sevensize CLEAN} algorithm, the blue triangles show the positions of the sets of pixels. Bottom panel: zoom on one of the sets of pixels.}
	\label{CLEANex}
\end{figure*}

Fig.~\ref{CLEANex} shows an example of this procedure.

It is then easy, in the frame of a Monte Carlo analysis based on this concept, to measure an astrometric precision which is defined as the root mean square (RMS) of the difference between the two sets of coordinates:

\begin{eqnarray}
	\label{sigma}
\sigma_{x} &=& \mathrm{RMS}(x_{i}^{\mathrm{model}}-x_{i}^{\mathrm{deconv}}) \\ \nonumber
\sigma_{y} &=& \mathrm{RMS}(y_{i}^{\mathrm{model}}-y_{i}^{\mathrm{deconv}}) \nonumber
\end{eqnarray}

All these simulated data are computed using the following parameters. 

We consider one night of observation chosen on 2010 April 15. Choosing the $15^{\mathrm{th}}$ day of the month is purely arbitrary. However, the choice of the month has an impact on the total time during which the Galactic centre (GC) is observable. April is quite a typical choice as the minimum (non-zero) observable length is in February and the maximum in June. During the chosen night, the GC is observable for a little more than 5 hours (as a comparison, it would be observable for 9 hours in June). 

One elementary integration block lasts 100~s. Two integration times are used: 5~h (i.e., the whole night), or a single 100~s block. When the observation is performed during 5~h, the observing procedure is chosen as follows. Integration on the scientific target is performed during 1~h, divided into elementary integration blocks of 100~s, then 30~min are spent for calibration, and so on during the night. The Earth rotation during one elementary block is supposed to be negligible because the trace in the u-v plane during the 100s of integration is perfectly linear and can be replaced by its average. Six baselines are used (with the four UTs) and five wavelengths (2, 2.1, 2.2, 2.3 and 2.4 $\mu \mathrm{m}$). 

\begin{figure*}
\centering
	\includegraphics[width=5cm,height=5cm]{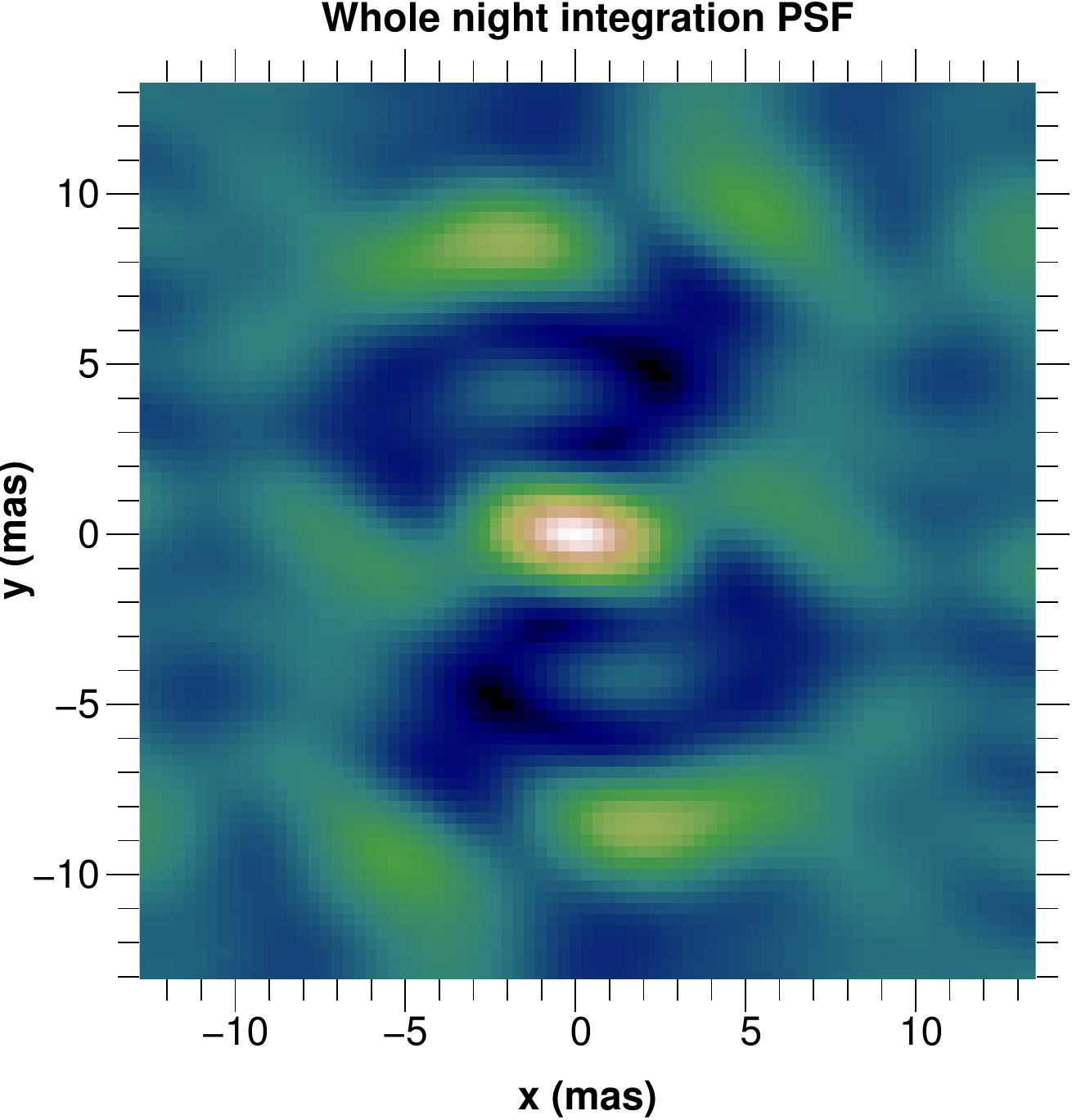}
	\includegraphics[width=5cm,height=5cm]{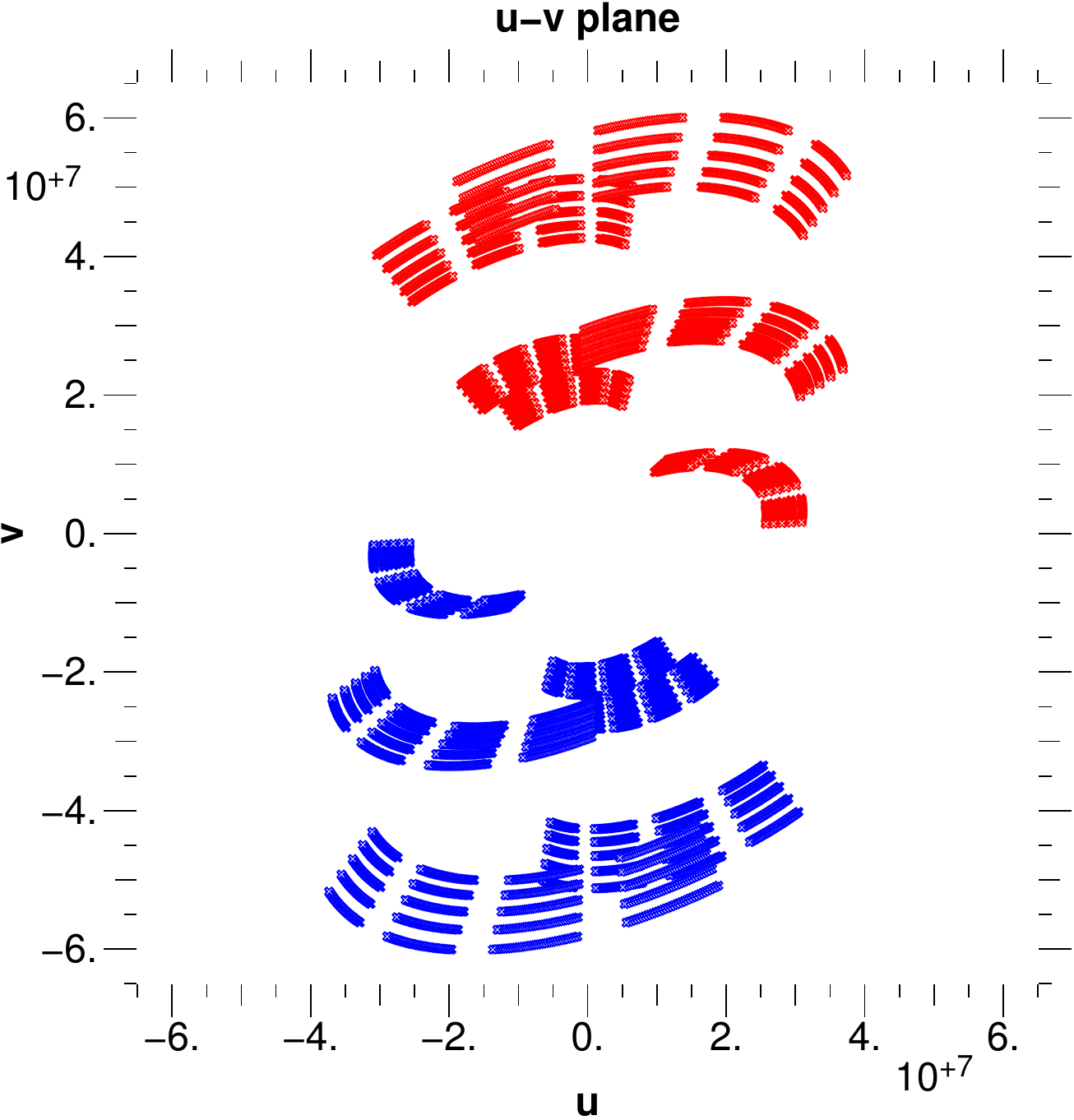}
	\caption{PSF central peak and u-v plane}
	\label{psf}
\end{figure*}

Fig.~\ref{psf} shows the u-v plane coverage obtained after 5~h of observation. The u-v plane has been rotated so that the $x$ and $y$ axes in the image plane are aligned with the big and short axes of symmetry of the PSF respectively.

The sampling (number of pixels per mas) is chosen in order to keep computing time reasonable, and to obtain the best possible precision. When the whole night is used to integrate, a 3 pixels per mas sampling is chosen. Finer sampling would demand too long a computing time in the Monte Carlo approach. When the integration time is 100~s, a 10 pixels per mas sampling is reasonable in terms of computing time, and can allow getting a finer precision than with 3 pixels per mas. However, this finer sampling is only of interest when a single star is in the field. If more than one star is present, the PSF of the sources overlap and contaminate each other, leading to the apparition of spurious secondary peaks. In these cases, the 10 pixels per mas sampling is no longer interesting. Thus, a 3 pixels per mas sampling is used in order to gain computing time.

It must be noticed that these samplings fulfill Shannon's criterium \citep{shannon49}, as the maximal frequency of the signal is $\frac{\lambda_{\rm{min}}}{B_{\rm{max}}} \approx 1.5$ mas.

\section{\textit{GRAVITY} astrometric precision in reconstructed images}
\label{precnoflare}

In this section, the astrometric precision in reconstructed images is computed when one to three stars are present in the field. Magnitudes are varied in the range $12 \leq m_{\mathrm{K}} \leq 17$. The integration time will be equal to 5~h (a configuration which can be used in practice to image a field of nearly motionless stars), or to a single 100~s block (which can be used to image a quickly evolving field, such as a flare).

For each Monte Carlo simulation made of $N$ different runs, run number $1\leqslant i\leqslant N$ gives a value of the differences $d_{i}^{x}$ and $d_{i}^{y}$ between deconvolved and model coordinates:

\begin{eqnarray}
	\label{dxy}
d_{i}^{x} &=& x_{i}^{\mathrm{model}}-x_{i}^{\mathrm{deconv}},\\ \nonumber
d_{i}^{y} &=& y_{i}^{\mathrm{model}}-y_{i}^{\mathrm{deconv}}.\nonumber
\end{eqnarray}

These quantities are in turn used to compute the RMS defined in equation~(\ref{sigma}).

The one-star case is of purely theoretical interest in order to determine the best possible precision in reconstructed images. Indeed, when the field of view contains only one point-like source, the standard two-field narrow angle astrometric method would provide better results because parametric reconstruction (i.e. fitting) is more efficient than non parametric methods (such as our imaging mode algorithm) as soon as the model is correct \citep[see][]{lebesnerais08}. However, we will show that in the 100~s integration case, the imaging mode results are very close to the fitting ones.

\begin{figure*}
\centering
	\includegraphics[width=5cm,height=5cm]{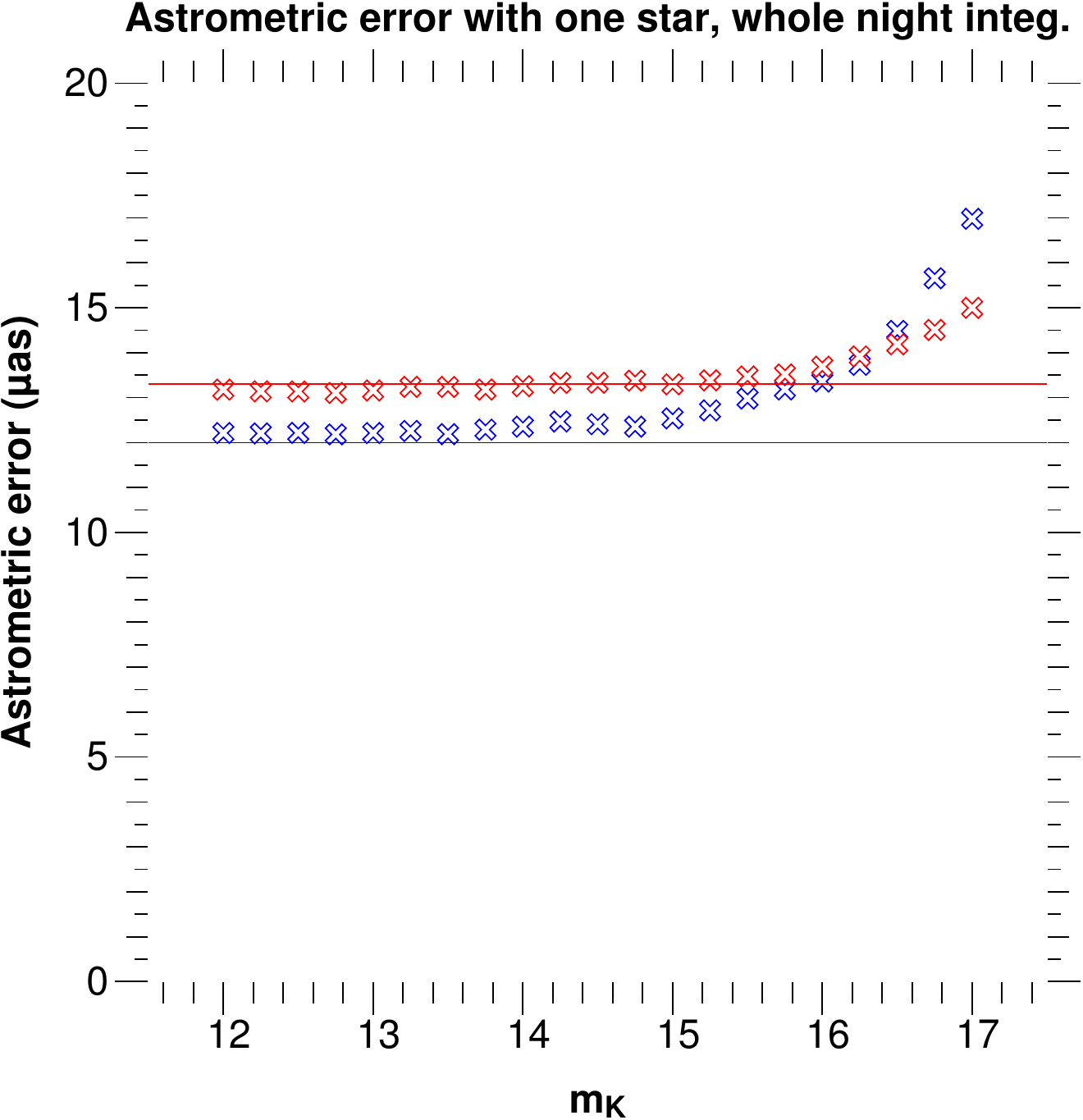}
	\includegraphics[width=5cm,height=5cm]{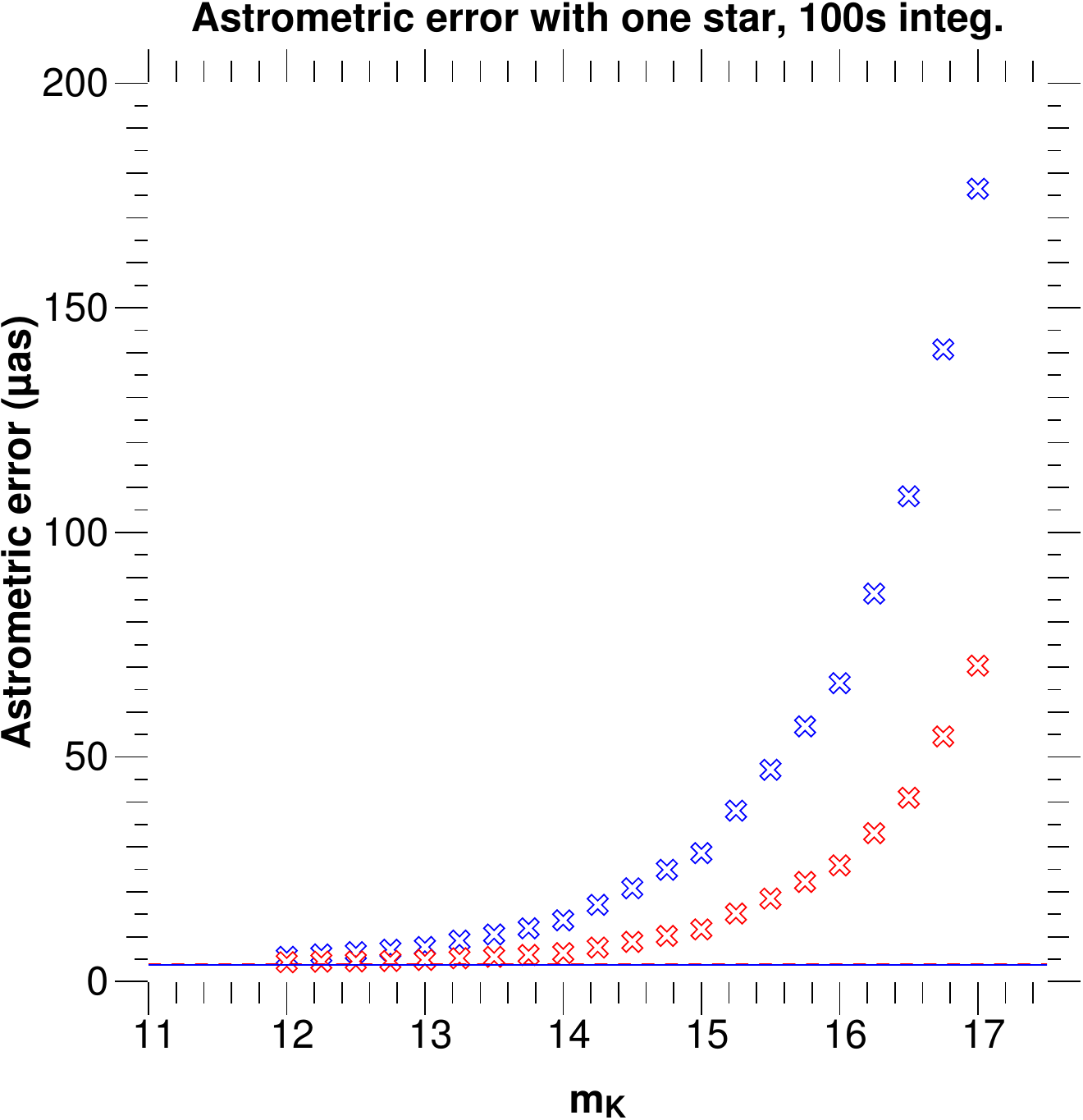}
	\caption{\textit{Imaging mode astrometric precision} in $\mu$as when \textit{one star} is present in the field of view as a function of the magnitude; in the $x$ direction (blue) and $y$ direction (red). The total integration time is equal to the whole night (left) or to 100 seconds (right). The solid horizontal lines show the level of precision without any noise added to the complex visibilities.}
	\label{1SXY}
\end{figure*}

\begin{figure*}
\centering
	\includegraphics[width=5cm,height=5cm]{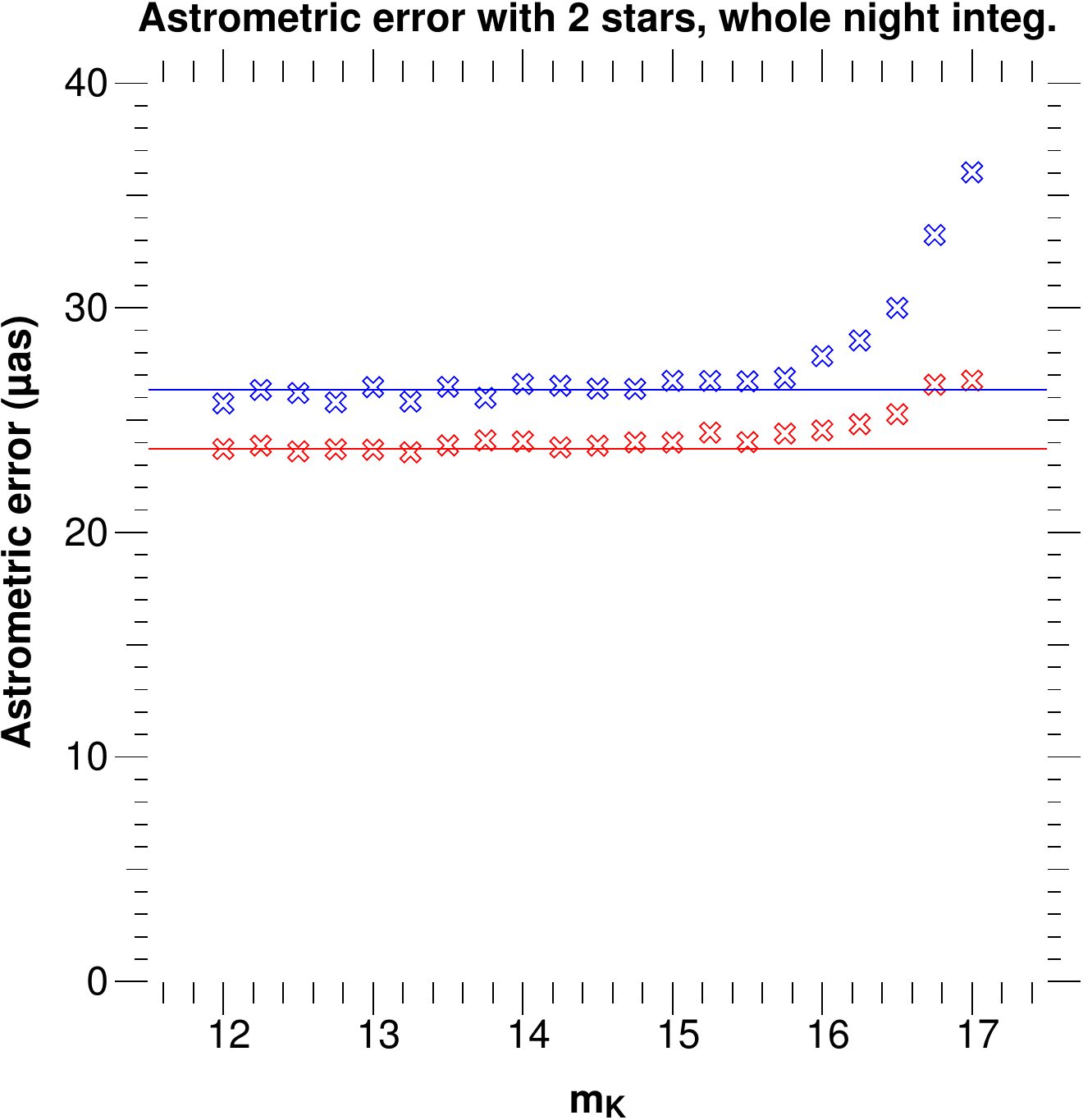}
	\includegraphics[width=5cm,height=5cm]{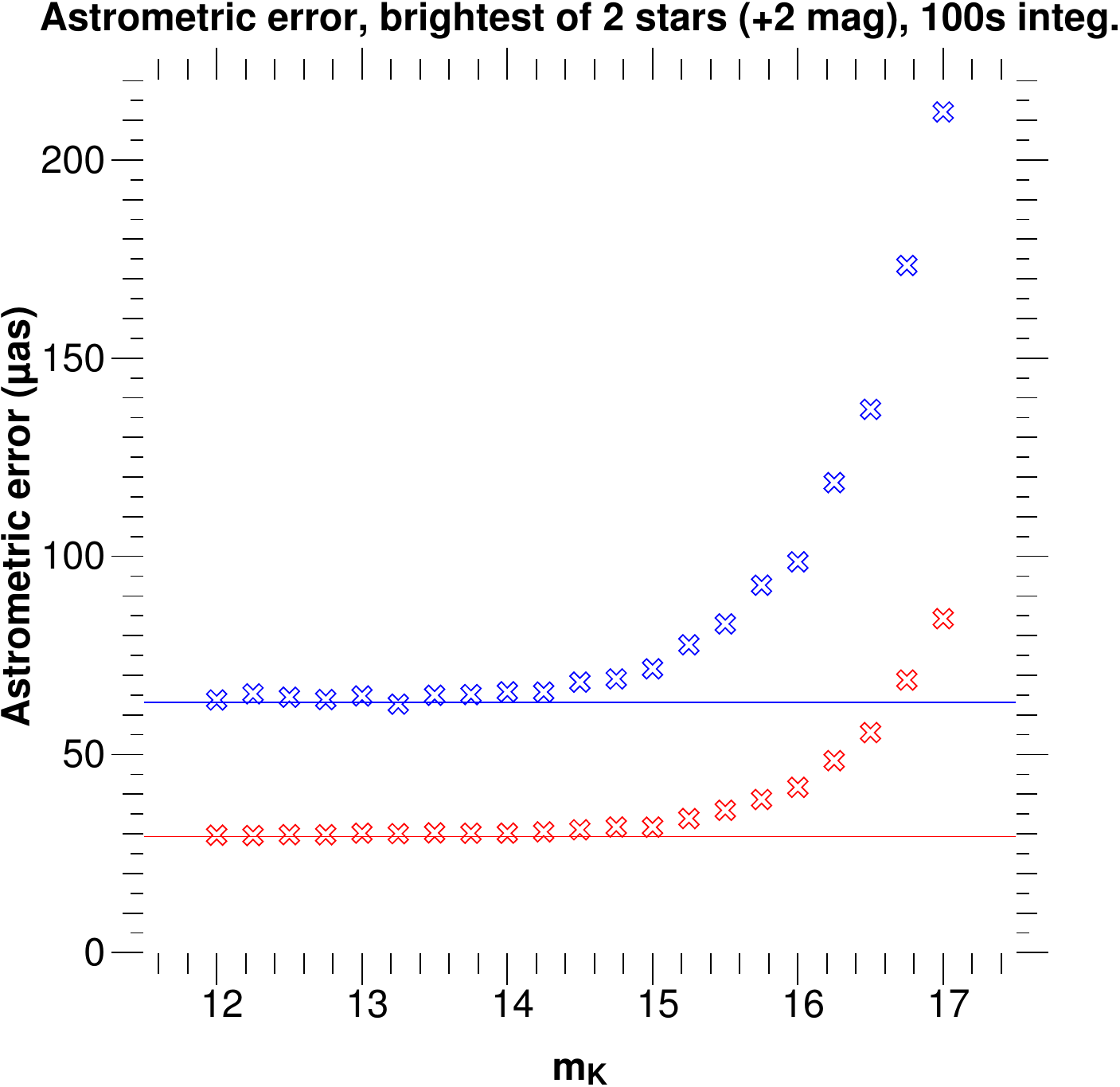}
	\caption{\textit{Imaging mode astrometric precision} in $\mu$as when \textit{two stars} are present in the field of view as a function of the integrated magnitude; in the $x$ direction (blue) and $y$ direction (red). The total integration time is equal to the whole night (left) or to 100 seconds (right). In this last case, only the brightest source is searched for, which is two magnitudes brighter. The solid horizontal lines show the level of precision without any noise added to the complex visibilities.}
	\label{2SXY}
\end{figure*}

\begin{figure*}
\centering
	\includegraphics[width=5cm,height=5cm]{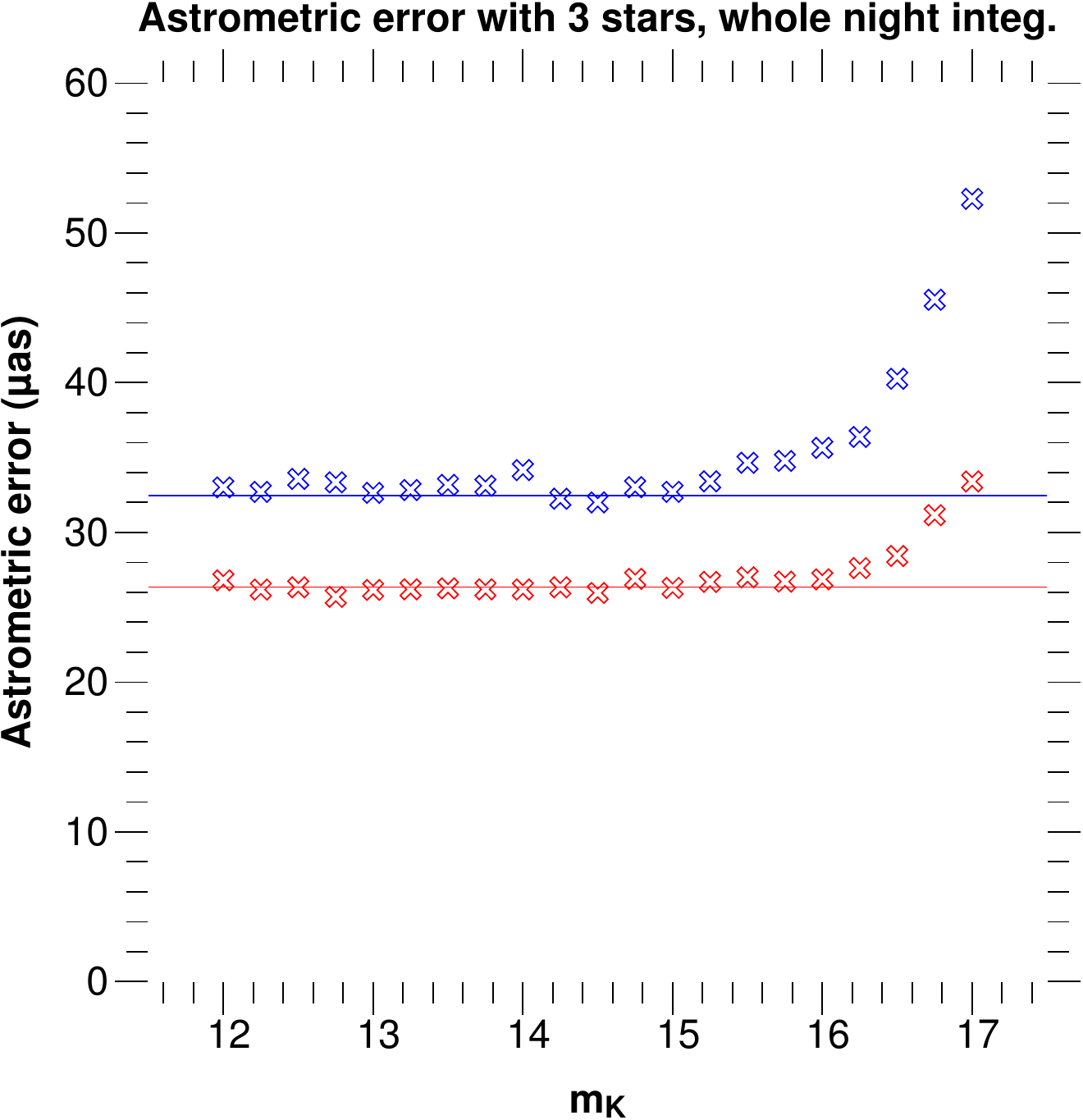}
	\includegraphics[width=5cm,height=5cm]{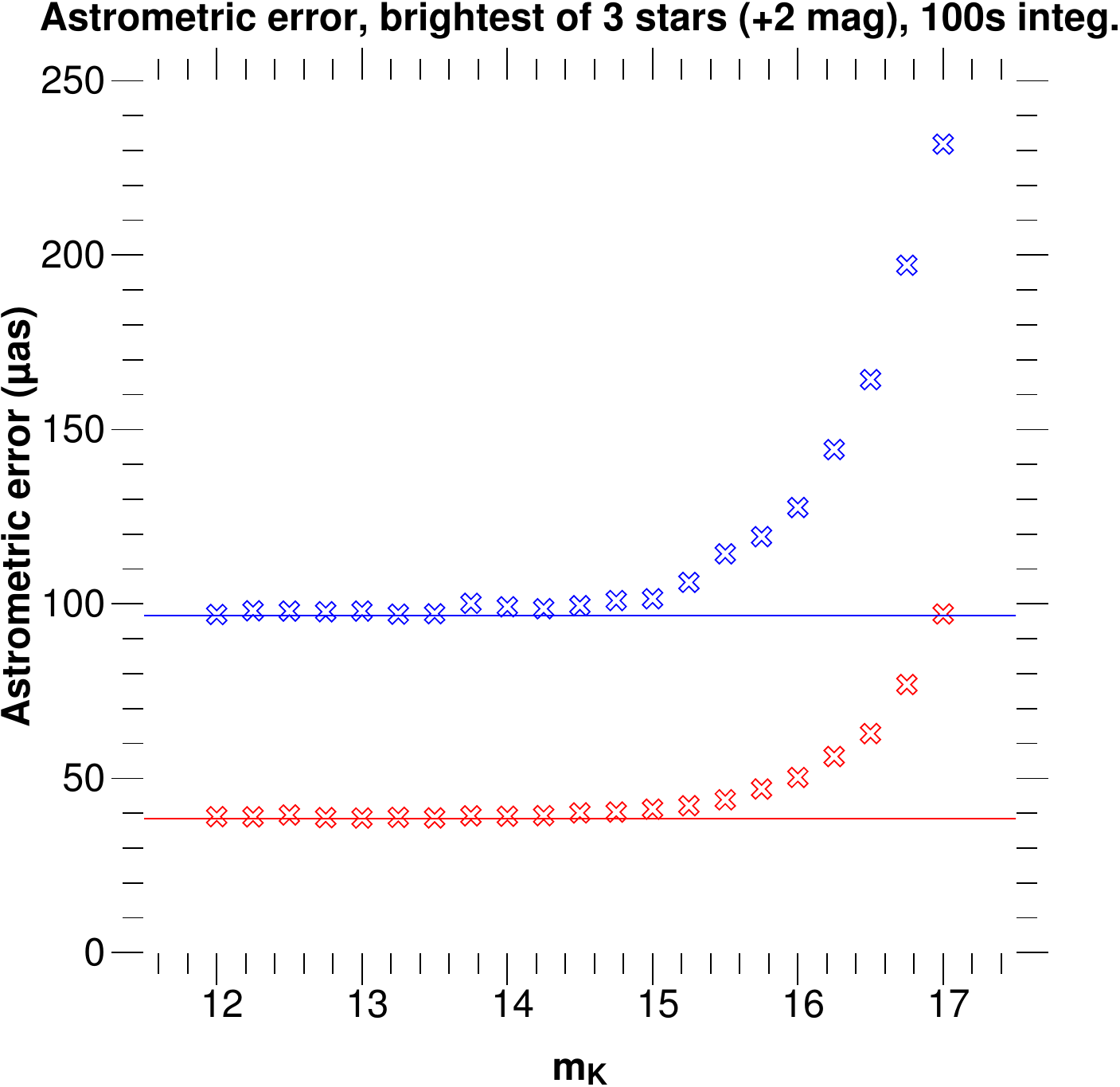}
	\caption{\textit{Imaging mode astrometric precision} in $\mu$as when \textit{three stars} are present in the field of view as a function of the integrated magnitude; in the $x$ direction (blue) and $y$ direction (red). The total integration time is equal to the whole night (left) or to 100 seconds (right). In this last case, only the brightest source is searched for, which is two magnitudes brighter. The solid horizontal lines show the level of precision without any noise added to the complex visibilities.}
	\label{3SXY}
\end{figure*}

Figs.~\ref{1SXY},~\ref{2SXY} and~\ref{3SXY} show the resulting astrometric precisions for the two different integration times and in the two directions $x$ and $y$. 

\begin{figure*}
\centering
	\includegraphics[width=5cm,height=5cm]{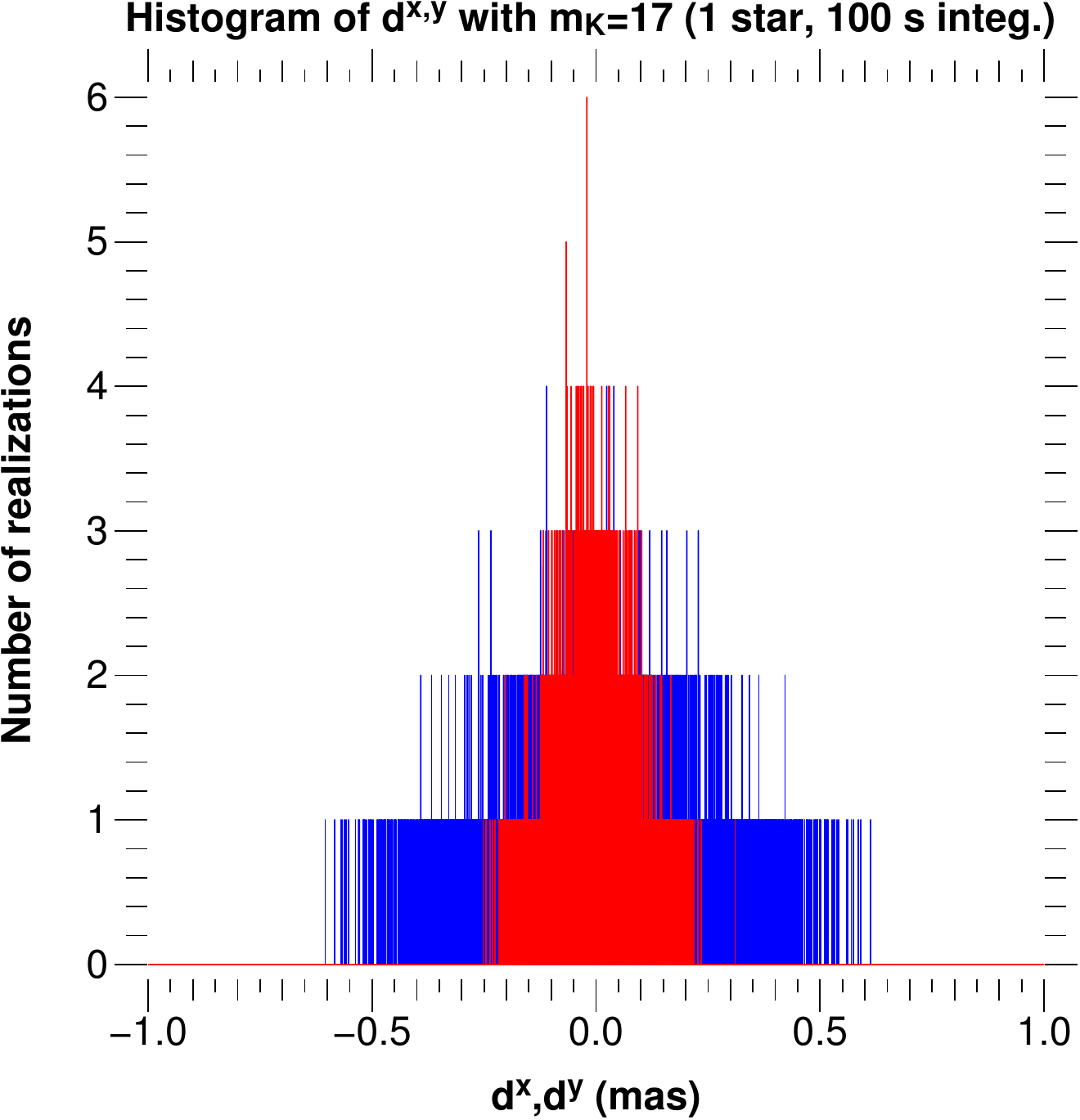}
	\includegraphics[width=5cm,height=5cm]{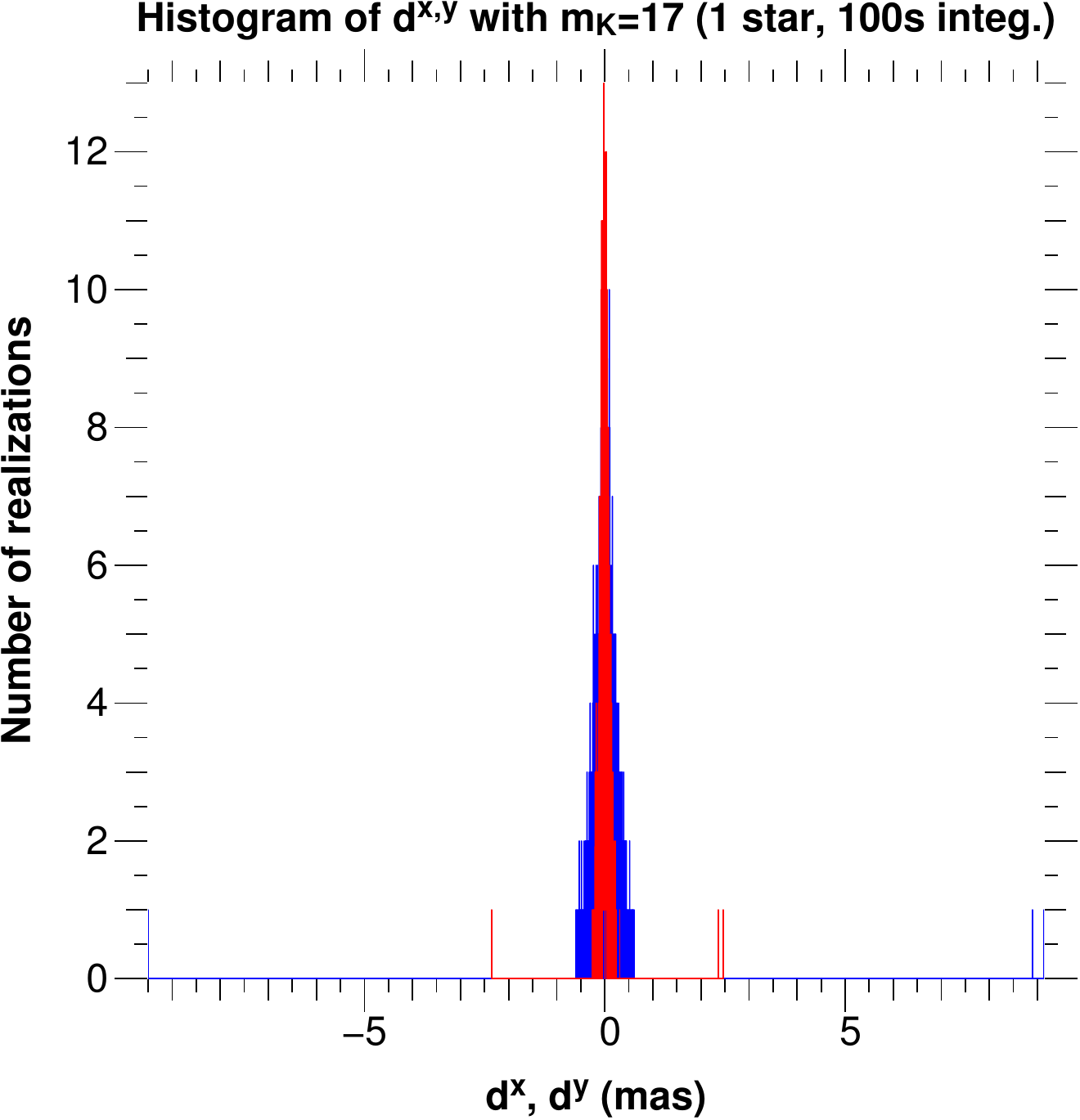}
	\caption{Histogram of the quantities $d^{x}$ (blue) and $d^{y}$ (red) defined in equation~(\ref{dxy}) with one $m_{\mathrm{K}}=17$ star in the field and 100~s of integration. Besides the gaussian shaped distribution (left), three realizations occur at values of $d^{x}$ and $d^{y}$ between 2 and 10 mas (right), thus disconnected from the broad distribution. These are `bad runs': due to a high level of noise in the image, the sharpest peak of the image is shifted from the true position of the source.}
	\label{Histo1SXY}
\end{figure*}

\subsection{General remarks}
\label{gene}

The astrometric error as a function of magnitude can be described with two main components: a plateau up to a cut-off magnitude and a steep increase of the error at fainter magnitudes. The plateau corresponds to a regime in which the astrometric accuracy is dominated by numerical errors of the {\sevensize CLEAN} procedure as shown by the horizontal solid lines in Figs.~\ref{1SXY},~\ref{2SXY} and~\ref{3SXY}. These solid lines show the level of precision obtained with noiseless data, i.e. when the only error is due to {\sevensize CLEAN} (which will be labeled `{\sevensize CLEAN} error' in the following). 

The steep increase of the error at fainter magnitudes is due to the increasing importance of the detection noise. 

The value of the plateau-limited precision is directly linked to the sampling. For instance in Fig.~\ref{1SXY}, the plateau is at $12 \, \mu $as for the 5~h case (3 pixels per mas sampling), and $3.7 \, \mu $as in the 100~s one (10 pixels per mas sampling). Would the sampling be set to 3 pixels per mas in the 100~s case, the plateau value would rise to approximately $12 \, \mu $as, as for the 5~h case.

Moreover, the better precision obtained in the $y$ direction is due to the shape of the PSF the axes of which are not equal: the $y$ axis being shorter, the precision in this direction is better.

We have tried to use one of the sources in the field of view as a reference in order to determine the relative astrometric precisions on the other two stars. However, even with a purely correlated error, the simulations show that there is no gain in using a position reference inside the field of view: the reconstruction errors in the coordinates retrieved are too high and spoil the relative precisions.

\subsection{Skipped Monte Carlo runs}

For some simulations, when the noise level is big enough, the {\sevensize CLEAN} algorithm can behave badly and generate output coordinates very different from the model ones. These `bad runs' are very rare and can be neglected, not to pollute the resulting precision. Indeed, only a very small fraction of such `bad runs' can drive the precision from a few tens of microarcseconds to the arcsecond.

In order to decrease the number of `bad runs', the standard procedure described in Section~\ref{method} is adjusted for some cases. 

When the integration time is equal to 5~h, `bad runs' occur when more than one star are present in the field and when they are very close to each other. Thus in these cases, the stars are forced to be further from each other than 5 mas, the PSF central peak's characteristic dimension. 

When the integration time is equal to 100~s, there are two main ways to decrease the number of `bad runs'. First,  the model coordinates $(x^{\mathrm{model}},y^{\mathrm{model}})$ can be restrained to be picked in a disc of radius 35 mas (i.e., the gaussian field of view of \textit{GRAVITY}) instead of being picked in the whole 100 mas square box. Second, the {\sevensize CLEAN} algorithm itself can be somewhat modified in order to prevent pathological behaviors. The standard working of the {\sevensize CLEAN} algorithm is the following. 

Let us consider an observed image and a `{\sevensize CLEAN}ed' image, initially identically zero.
\begin{enumerate}
	\item Step 1:  The maximum value of intensity in the observed image is located. 
	\item Step 2: The same pixel in the {\sevensize CLEAN}ed image is added a given fixed value of intensity. 
	\item Step 3: The corresponding amount of the PSF, multiplied by a loop-gain factor, is subtracted at the maximum location on the observed image. 
	\item Back to Step 1.
\end{enumerate}

The stopping condition is that the new maximum of the observed image becomes less than a given fraction of its initial maximum:

\[
\mathrm{max}_{\mathrm{new}} \leqslant \frac{\mathrm{max}_{\mathrm{init}}}{D}
\]
where $D$ is a chosen parameter, the value of which is taken between a few and 10. 

Problems can arise when both the noise and $D$ are high. Then, by subtracting many PSF on the true maximum of the image, some spurious secondary maxima may become sharper in the {\sevensize CLEAN}ed image, leading to a completely wrong retrieved value of the coordinates. An easy way to prevent this is to verify at the end of the {\sevensize CLEAN} algorithm that the {\sevensize CLEAN}ed image maximum is close to the initial maximum of the observed image (of course this only insures that the brightest star is not a fake in the {\sevensize CLEAN}ed image, but we will see in the following that only the brightest star can be retrieved when many sources are present with an integration time of 100~s). If these two maxima do not coincide, then the {\sevensize CLEAN} algorithm is restarted with a smaller value of $D$.

These modifications of the standard procedure defined in Section~\ref{method} are always used (for an integration time of 100~s) when more than one star are present in the field, and even with a single star, when $m_{\mathrm{K}} \geqslant 16.5$.

However, despite all these precautions to prevent `bad runs', some may remain: if, due to noise, the maximum of the observed image is shifted from the true star position, there is no way to find the source's position with {\sevensize CLEAN}. These pathological cases will be removed for the final precision computation. In the following sections, it is always specified whether `bad runs' were removed, and the percentage of these skipped runs is provided.

\subsection{One-star case}

Here, `bad runs' only occur when the integration time is 100~s and for $m_{\mathrm{K}}=17$. Fig.~\ref{Histo1SXY} shows the histogram of the differences $d_{i}^{x}$ and $d_{i}^{y}$ defined by equation~(\ref{dxy}) in this case. It appears that, besides a peak located near zero, some rare events are present at quite higher values. In order not to pollute the result, the final precision is computed without using the runs that end up with $d_{i}^{x}$ or $d_{i}^{y}$ greater than 2 mas (the first polluting peak in Fig.~\ref{Histo1SXY} is indeed located a little higher than 2 mas). The proportion of runs skipped in this procedure is of 0.03 per cent.

The {\sevensize CLEAN} algorithm is used here in its optimal configuration with only one point source. In this specific case, its result is nearly independent of the shape of the PSF: the {\sevensize CLEAN} error is weakly dependent on the direction and on the integration time, it is mainly dictated by the sampling, as stressed in Section~\ref{gene}.  

Of course, the precision cannot be ever more improved even if the sampling is refined: it will always remain greater than the theoretical limiting precision.

The best precision obtained is by far better than the pixel size. For instance in the 5~h integration case, the pixel size is $0.33$ mas so the precision is better than the pixel size by a factor of $25$. This is due to the fact that the barycentre of the illuminated pixels is calculated, which makes it possible to obtain sub-pixel precision.

It must be emphasized that the best precision obtained in the 100~s case, 4 $\mu$as, is very close to the theoretical limit: this is a powerful demonstration of the interest of the imaging mode. Moreover, this theoretical precision can be obtained provided the sampling is high enough. With 150 pixels per mas, the pure imaging mode precision in the $y$ direction reaches $1.6 \, \mu$as. However, this is done at the expense of computing time.

It is interesting to check whether one reaches the theoretical limit error when using the pure imaging mode results, with a 10 pixels per mas sampling, to bootstrap a fitting procedure. The complex visibilities are fitted to the simulated data, the only parameters being the $x$ and $y$ angular position of the source on the sky. The intensity is not fitted, it is supposed to be measured with a 10 per cent error by the adaptive optics of the instrument.

The imaging mode is used to determine a first guess of the position of the star in the field. More precisely, the first guess is chosen by sampling a box centered on the imaging mode retrieved position, its dimension being of the order of the imaging mode precision at the considered magnitude. The iteration is stopped as soon as the $\chi^{2}$ corresponding to a position in the box is between 0.5 and 1.5. Then the fit is performed with Yorick's \verb|lmfit| routine. This procedure allows to have access to a precision of $1.6 \, \mu$as when integrating 100~s on a $m_{\rm{K}}=12$ source. This coincides with the theoretical value given in Section~\ref{imorfit}. 

\subsection{Two-star case}

With two stars in the field and when integrating during the whole night, Fig.~\ref{2SXY} shows that the astrometric precision is of the order of the Schwarzschild ISCO angular radius (30 $\mu$as) in both directions and for all magnitudes.

When the integration time is 100~s, histograms similar to Fig.~\ref{Histo1SXY} are too noisy to be able to determine a rejecting criterion (i.e. a limiting highest acceptable value for $d_{i}^{x}$ and $d_{i}^{y}$). Things become better only if one restricts to searching the brightest source, which is moreover supposed to be two magnitudes brighter than the second one. The other fainter star is not searched for by {\sevensize CLEAN}. This case is still interesting because this configuration can occur at the galactic centre if a bright flaring source is observed when a faint star is also present in the field: only the flare's position would be of interest. Fig.~\ref{2SXY} shows that the precision in the best direction for\footnote{Here $m_{\mathrm{K}}$ is the magnitude integrated over the field of view, not the magnitude of the brightest star only.} $m_{\mathrm{K}} \leqslant 15$ is here too of the order of 30 $\mu$as. For all these 100~s results, the percentage of skipped runs is always zero, except for $m_{\mathrm{K}} \geqslant 16.75$ where it remains less than 0.1 per cent.

\subsection{Three-star case}

Fig.~\ref{3SXY} shows that when integrating the whole night, the precision in both directions is still of the order of the Schwarzschild ISCO for magnitudes higher than $m_{\mathrm{K}}=16$.

When the integration time is 100~s, and when searching only for the brightest source, the precision in the best direction for magnitudes higher than $m_{\mathrm{K}}=15.5$ is of the order of 40 $\mu$as, somewhat higher than the ISCO. The percentage of skipped runs is always less than 0.5 per cent.

\vspace{0.5cm}

The main numerical results of this section are summarized in Table~\ref{prectab}.

All these results are very encouraging as far as the GR-related science cases of \textit{GRAVITY} are concerned: the immediate vicinity of the black hole is within reach of the instrument even if the source is quite complex.

\begin{table*}
\centering 
\begin{minipage}{130mm}
\caption{Imaging mode best precision as a function of integration time and sampling (ppm = pixels per mas), and percentage of skipped `bad runs' (for 100~s integration). When more than one star are present and the integration time is 5~h, the stars are supposed to be more distant than the PSF characteristic dimension to avoid `bad runs'. When the integration time is of 100~s, only the brightest source is searched.} 
\label{prectab} 
\begin{tabular}{c c c c}
\hline
Nb of stars & 5~h & 100~s & \% skipped (for 100~s) \\
\hline
1 & 13 $\mu$as, 3 ppm & 4 $\mu$as, 10 ppm & $\leqslant$ 0.03\\ 
2 & 25 $\mu$as (not too close), 3 ppm & 30 $\mu$as (brightest), 3 ppm &  $\leqslant $ 0.1 \\
3 & 30 $\mu$as (not too close), 3 ppm &  40 $\mu$as (brightest), 3 ppm & $\leqslant$ 0.5 \\
\hline
\end{tabular}
\end{minipage}
\end{table*}

\section{Information retrieved when a variable source is in the field of view}
\label{precflare}

In this section, the field of view contains a variable source with maximum magnitudes $m_{\rm{K}}=$13, 14 or 15. These values are realistic for the galactic centre flares: we have seen in Section~\ref{intro} that the brightest flare observed to date has a maximum magnitude $m_{\rm{K}}=13.5$.

\begin{figure*}
\centering
	\includegraphics[width=5cm,height=5cm]{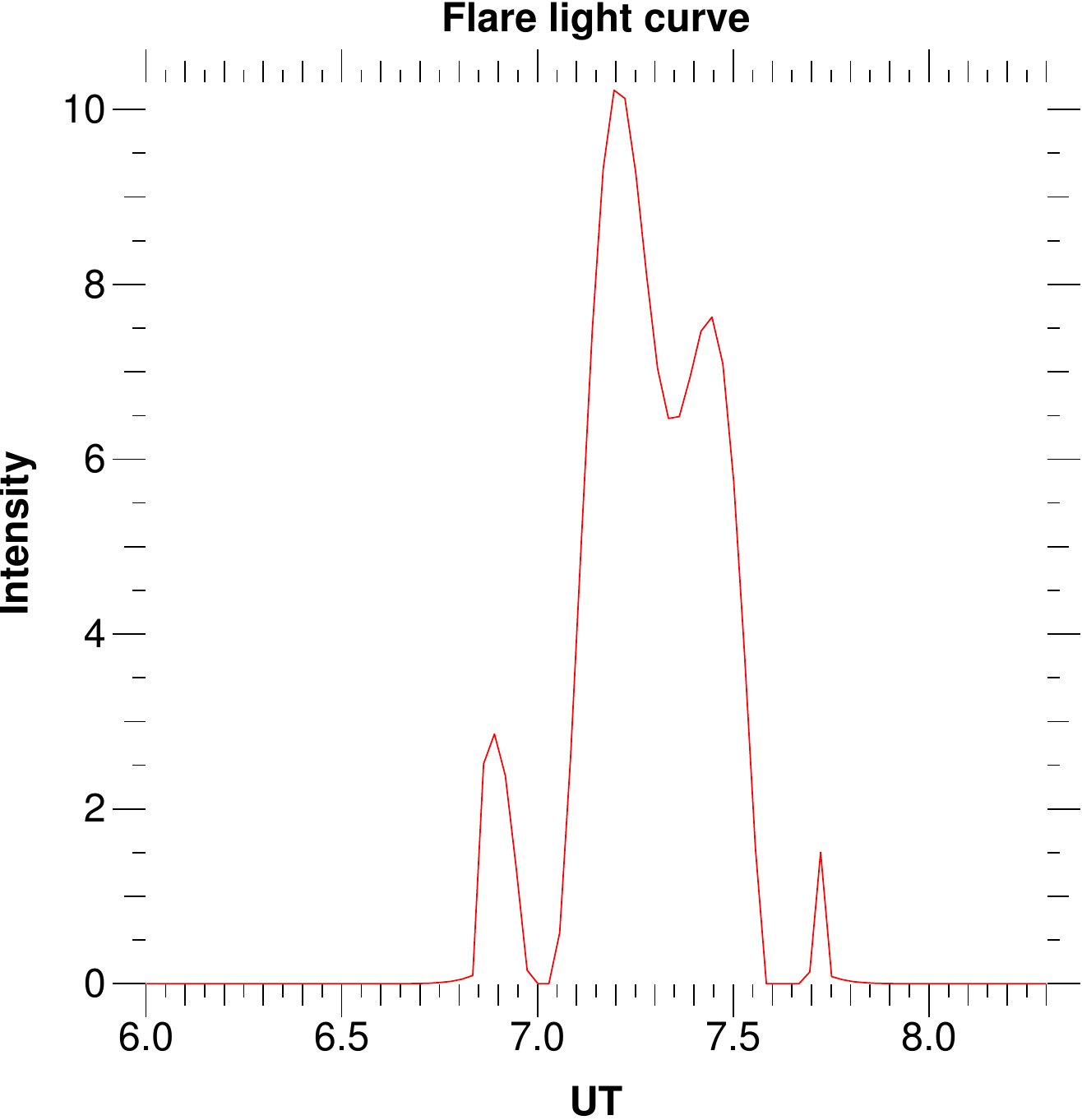}
	\caption{Light curve of the variable source. Intensities are in arbitrary units and the abscissa is graduated in hours, UT. }
	\label{flareI}
\end{figure*}

As for the brightness, it is modeled by the superimposition of a Gaussian and a sine curve (see Fig.~\ref{flareI}) to take into account the two characteristic time-scales of the flares mentioned in Section~\ref{intro}. It is assumed that the source light curve is measured with a 10 per cent accuracy by \textit{GRAVITY}'s adaptively corrected guiding cameras. This value of intensity will be used in the fitting procedure which will allow to obtain the source position.

Two kinds of simulations are performed. Simulations of the first kind are done with a non-moving source, with variable magnitude. Simulations of the second kind are done with a source both variable in position and magnitude. Its trajectory is circular, at a distance of $30 \, \mathrm{\mu as}$ from the centre, corresponding to the Schwarzschild ISCO of a black hole of $4.3 \, 10^{6} M_{\mathrm{\odot}}$  \citep[in this simple model, the lensing effects discussed e.g. in][are neglected]{hamaus09} .

In all simulations, an observation is assumed every 100~s during the whole night. Each observation gives a set of visibilities and phases. The period of the night in which the observed intensity is higher than half the maximum observed intensity is selected. An image is then reconstructed for each set of visibilities and phases in this interval. Fifteen such 100~s integration snapshots are thus obtained on both sides of the maximum of the flare.

The astrometric positions measured on the reconstructed images are found by resorting to pure imaging mode (with a 10 pixels per mas sampling), or by fitting the complex visibilities with a single-source model using the imaging mode retrieved positions as first guesses. 

One possible way to check whether a motion can be detected is to calculate the dispersion of the positions thus obtained, in both directions:

\begin{eqnarray}
	\label{dispersion}
\delta_{\mathrm{x}} & = & \sqrt{\frac{1}{14}\sum_{k=0}^{14} \left( x_{\rm{k}}-\bar{x}\right)^{2}},  \\ \nonumber
\delta_{\mathrm{y}} & = & \sqrt{\frac{1}{14}\sum_{k=0}^{14} \left( y_{\rm{k}}-\bar{y}\right)^{2}}\nonumber
\end{eqnarray}
where $x_{\rm{k}}$ and $y_{\rm{k}}$ are the retrieved positions of the flare in the $x$ and $y$ direction in each of the fifteen snapshots, and the barred quantities are the average position during the considered night.

\begin{figure*}
\centering
	\includegraphics[width=5cm,height=5cm]{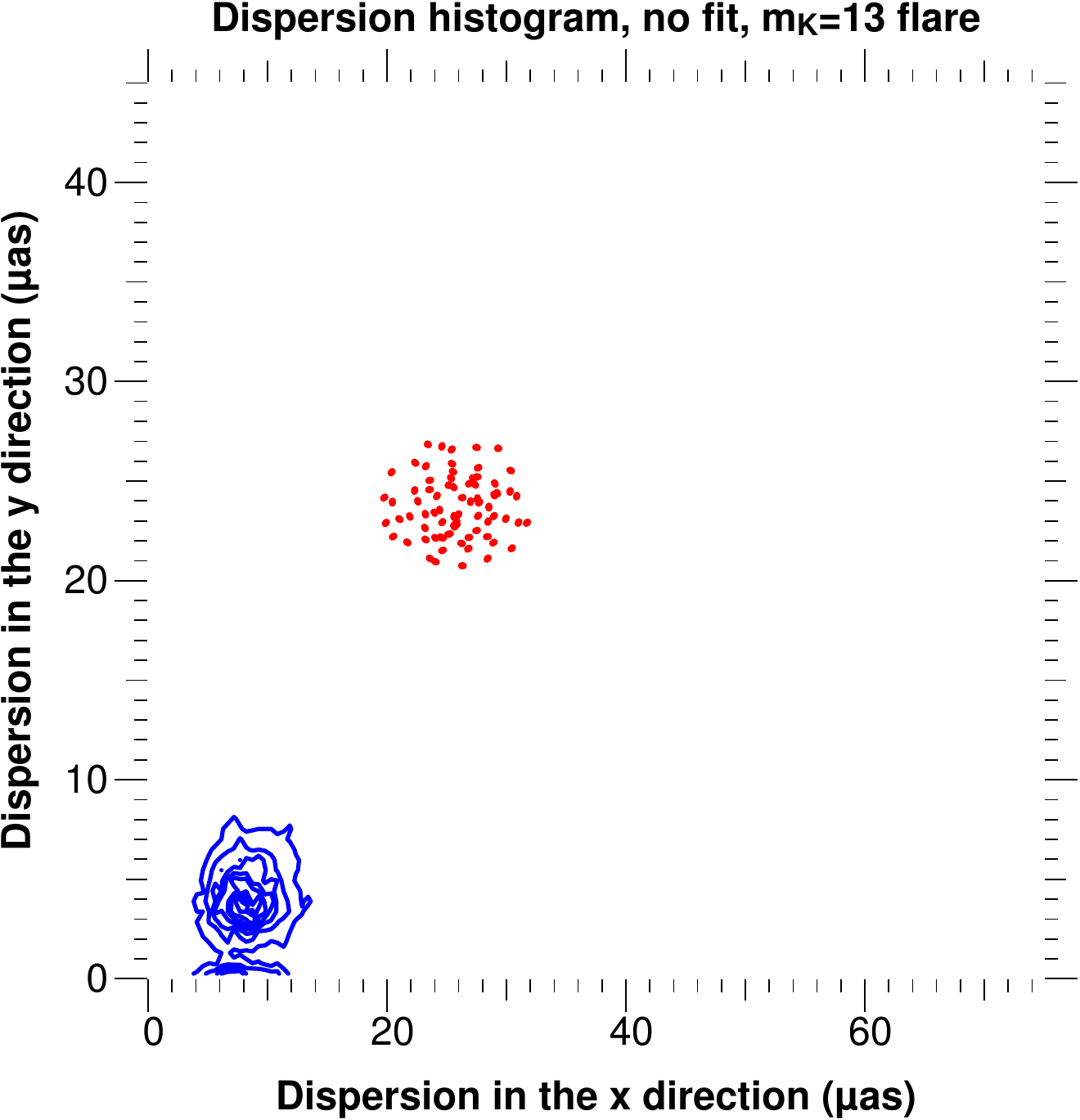}
	\includegraphics[width=5cm,height=5cm]{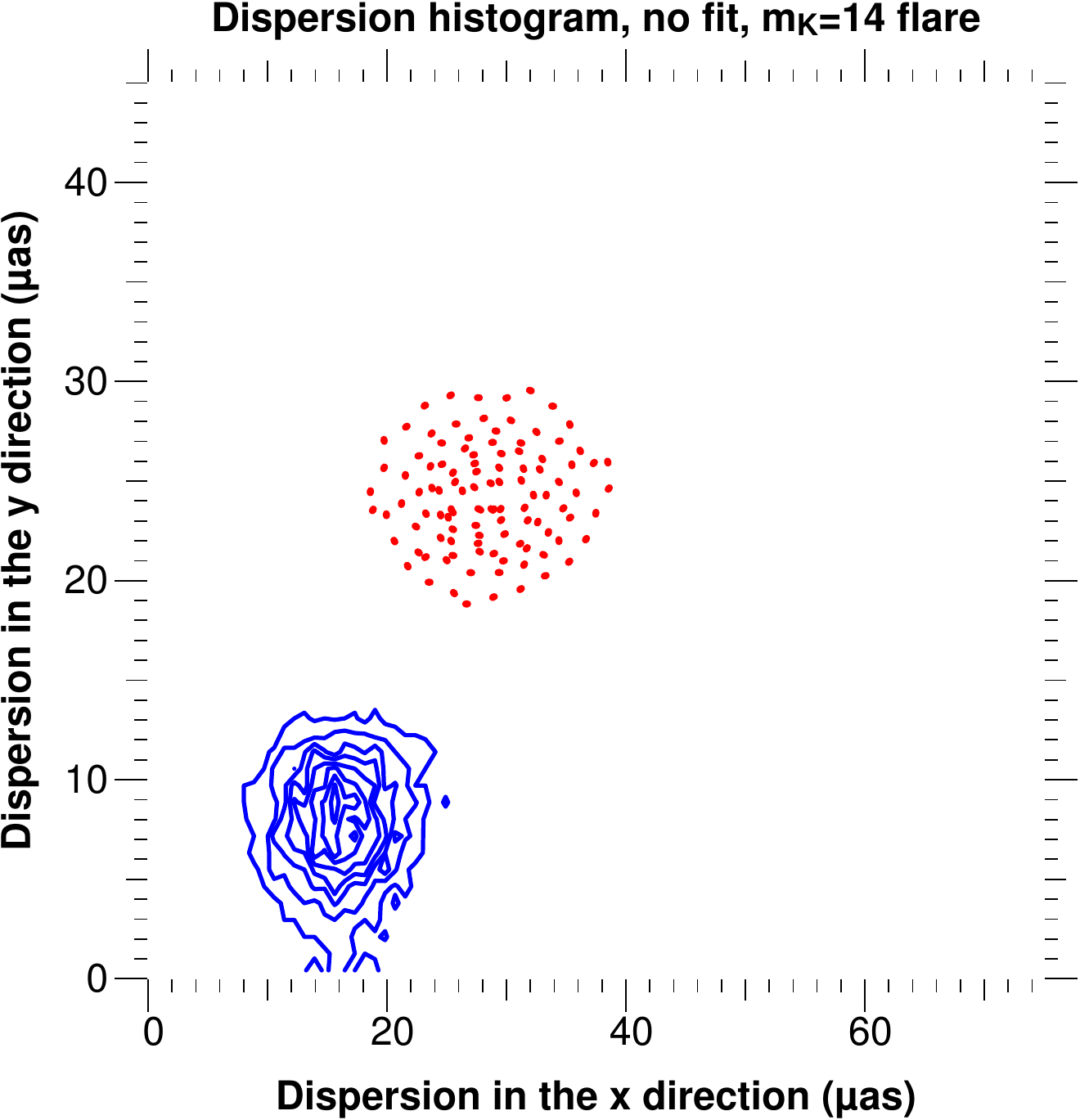}
	\includegraphics[width=5cm,height=5cm]{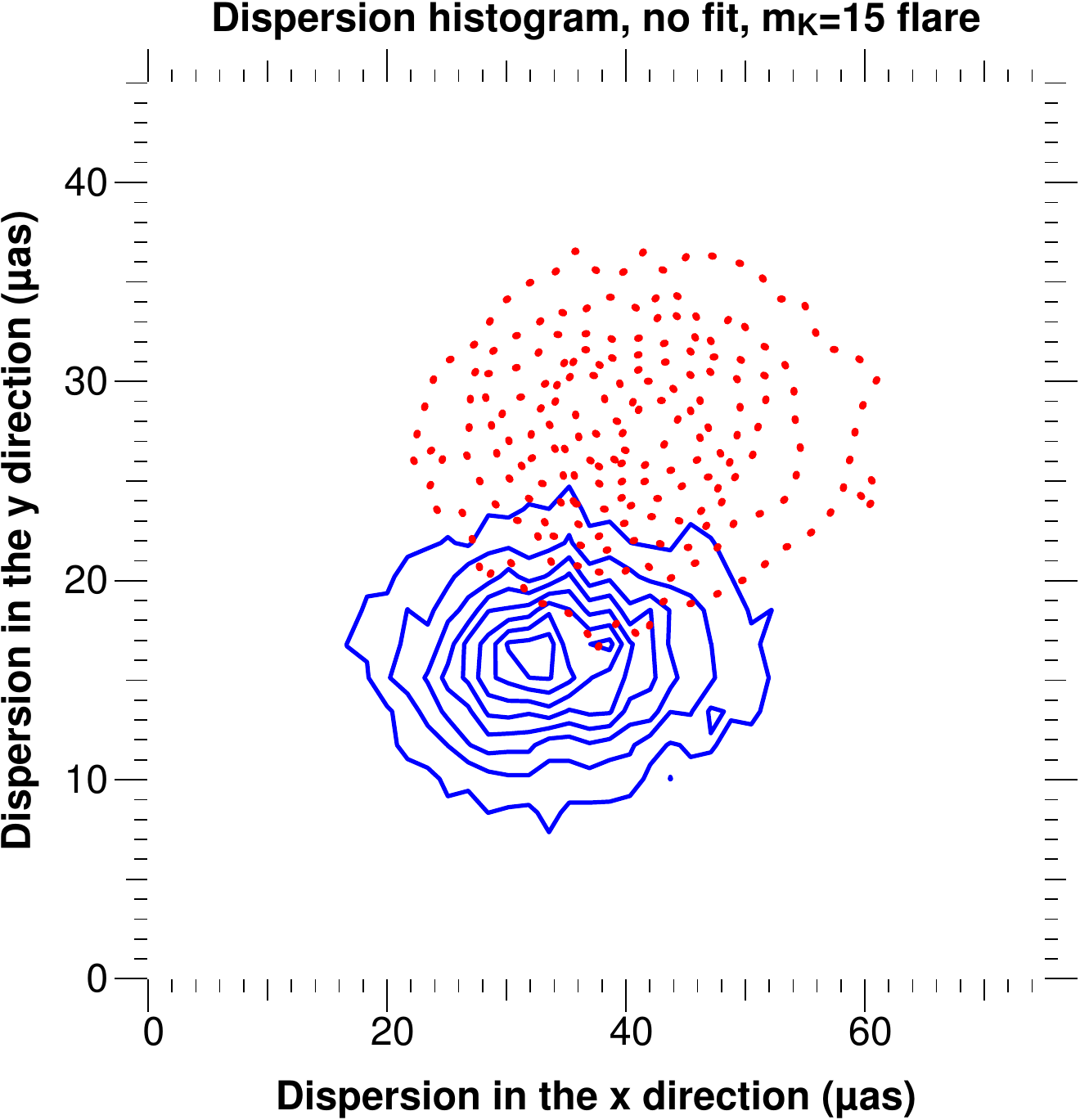}
	\caption{Isocontours of the histograms of the 2D dispersion for a moving flare (dotted red) and a non-moving one (solid blue); the maximum magnitude of the flare is 13, 14, 15 for upper-left, upper-right and bottom pictures. The retrieved positions are found in the imaging mode. The axes are graduated in $\mathrm{\mu as}$.}
	\label{histodispersion}
\end{figure*}

\begin{figure*}
\centering
	\includegraphics[width=5cm,height=5cm]{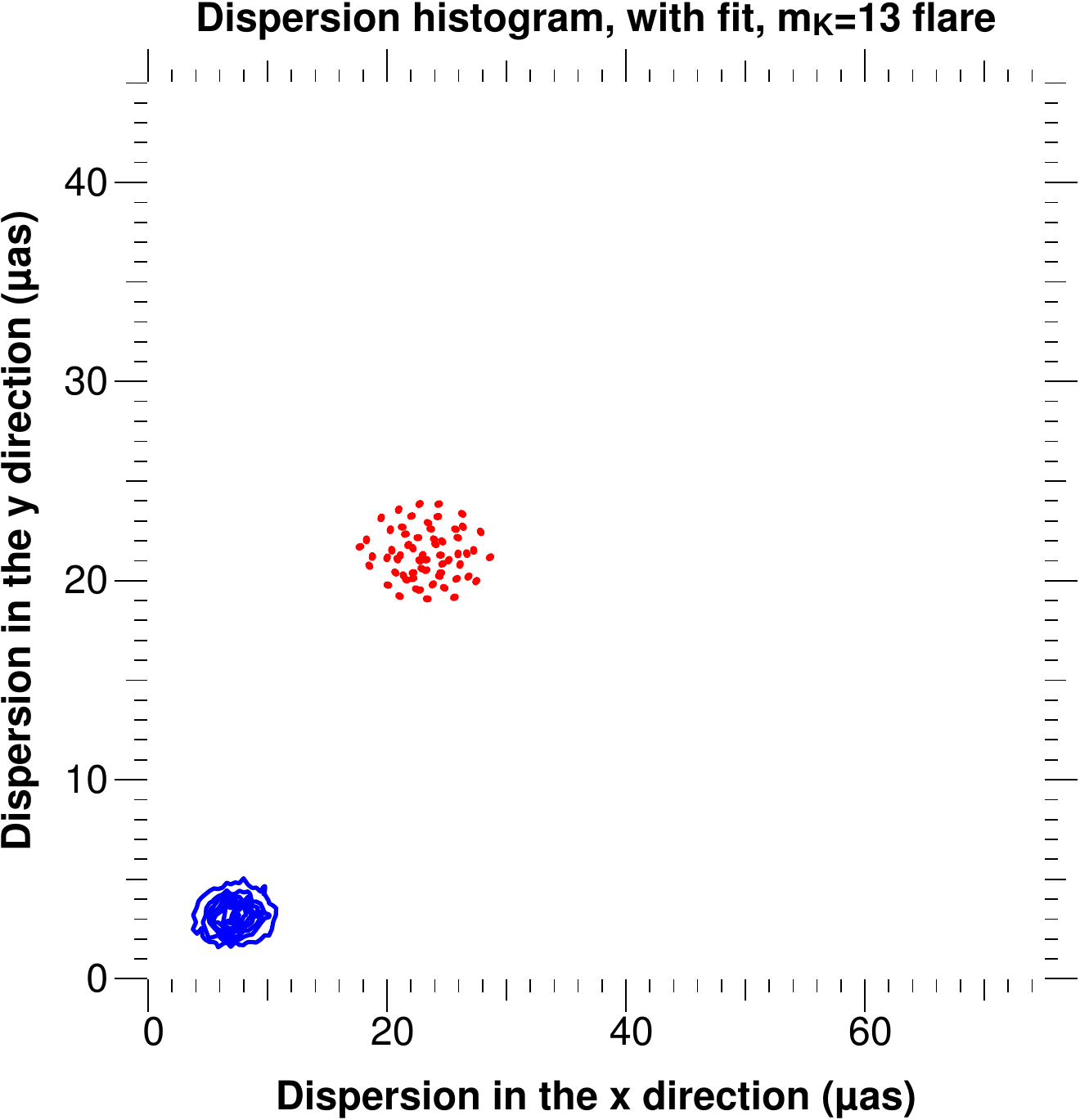}
	\includegraphics[width=5cm,height=5cm]{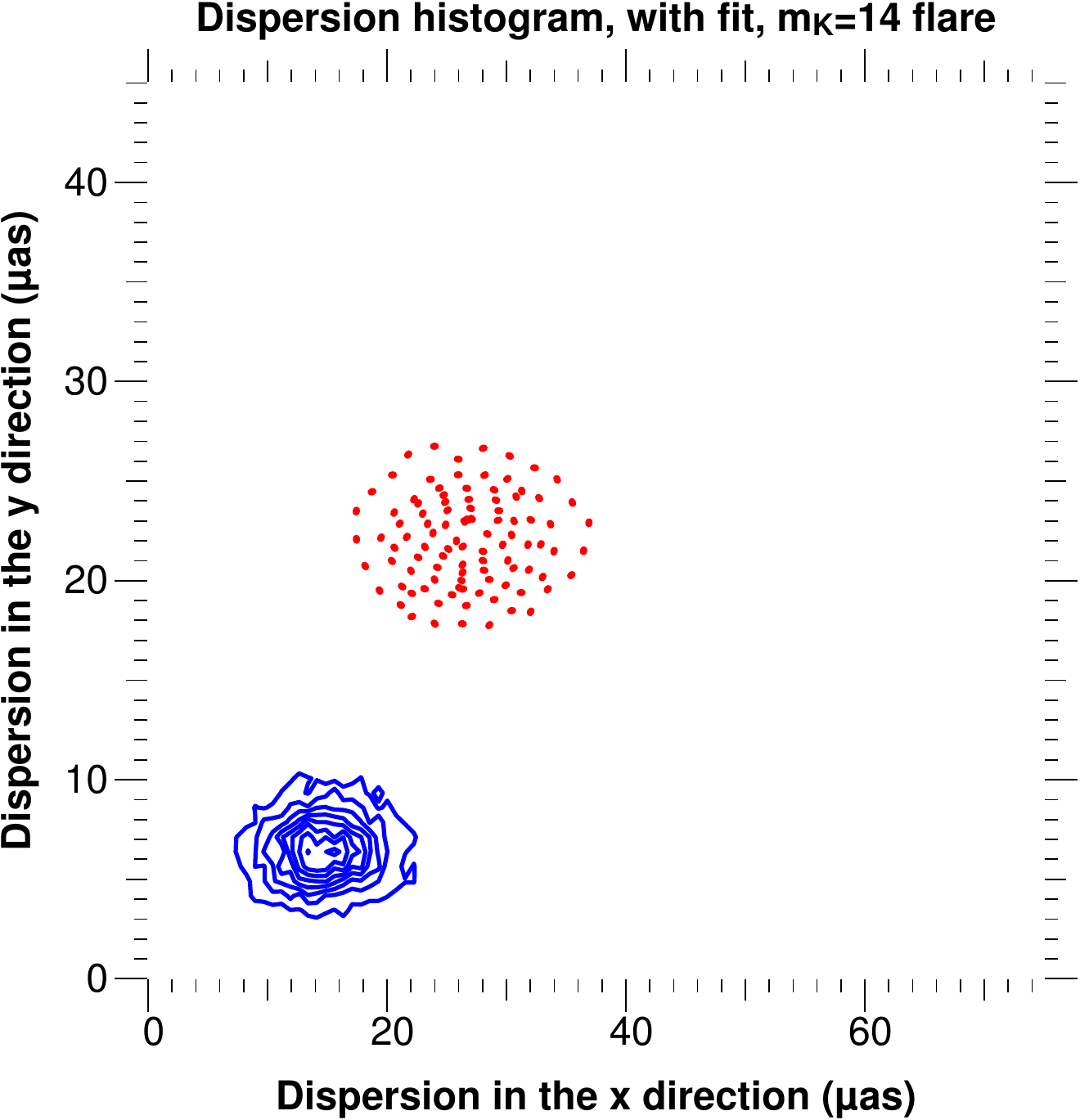}
	\includegraphics[width=5cm,height=5cm]{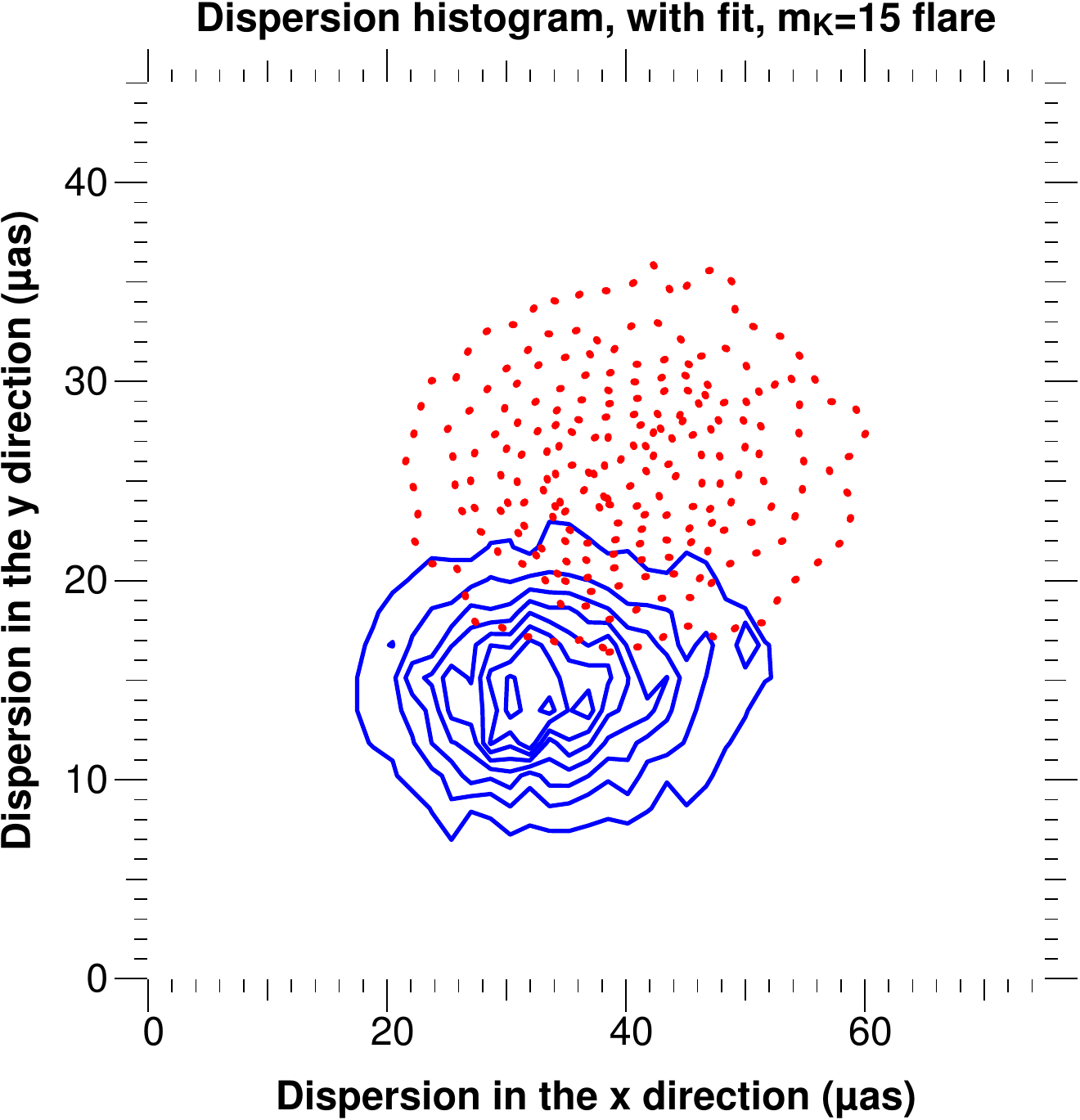}
	\caption{Same as Fig.~\ref{histodispersion}, the retrieved positions being found by resorting to fitting, the imaging mode results being used as a first guess. }
	\label{histodispersionfit}
\end{figure*}

By making a Monte Carlo analysis based on this scheme, the histogram of the 2D dispersion can be computed. Fig.~\ref{histodispersion} represents the isocontours of these 2D histograms in the pure imaging mode case. Fig.~\ref{histodispersionfit} gives the same results when a fitting procedure is used. In both figures, the blue contours are related to the simulations with a non-moving source, and the red contours are related to the simulations with a moving source.

In practice, one night of observation will give one value of the dispersion in both directions $(\delta_{\mathrm{x}}, \delta_{\mathrm{y}})$. The interest of these histograms is that if this point falls into one of the two sets of contours (blue or red), and if the two sets do not overlap or very slightly overlap, then it is possible to conclude whether the flare is moving or not. 

Of course, the model used here is very simple, because no relativistic effects are taken into account to model the flare motion, and the orbit is supposed to be seen face on. The effect of the inclination is not taken into account: it would shift the histograms. Moreover, if the orbit is seen nearly edge-on, relativistic effects will make it look very different, as described extensively in \citet{hamaus09}. Consequently, Figs.~\ref{histodispersion} and~\ref{histodispersionfit} must be taken as a first step towards a more realistic modeling, which will be done in future work.

It appears first on Fig.~\ref{histodispersion} and~\ref{histodispersionfit} that the histograms of non-moving sources (solid blue ones) are centered on a value of $(\delta_{\mathrm{x}}, \delta_{\mathrm{y}})$ which is close to the value of the astrometric precision (pure imaging mode precision, or `imaging mode + fit' precision, respectively) corresponding to one point source of the same magnitude. Indeed the pure imaging mode astrometric precisions for one star of magnitude $m_{\rm{K}} = 13, \,14, \, 15$ are in $\mu$as: $(\sigma_{\rm{x}}=7.7, \sigma_{\rm{y}}=4.8)$, $(\sigma_{\rm{x}}=13.6, \sigma_{\rm{y}}=6.3)$ and $(\sigma_{\rm{x}}=28.6, \sigma_{\rm{y}}=11.6)$. And the `imaging mode + fit' corresponding precisions are: $(\sigma_{\rm{x}}=6.7, \sigma_{\rm{y}}=2.6)$, $(\sigma_{\rm{x}}=12.8, \sigma_{\rm{y}}=5.0)$ and $(\sigma_{\rm{x}}=27.4, \sigma_{\rm{y}}=10.6)$. Moreover, it appears that the pure imaging mode results are as rich as the fitting ones: the histograms are very close to each other, which is not surprising as it has been shown in the preceding section that fitting does not allow to gain a lot in precision in the 100~s integration case.

\begin{figure*}
\centering
	\includegraphics[width=5cm,height=5cm]{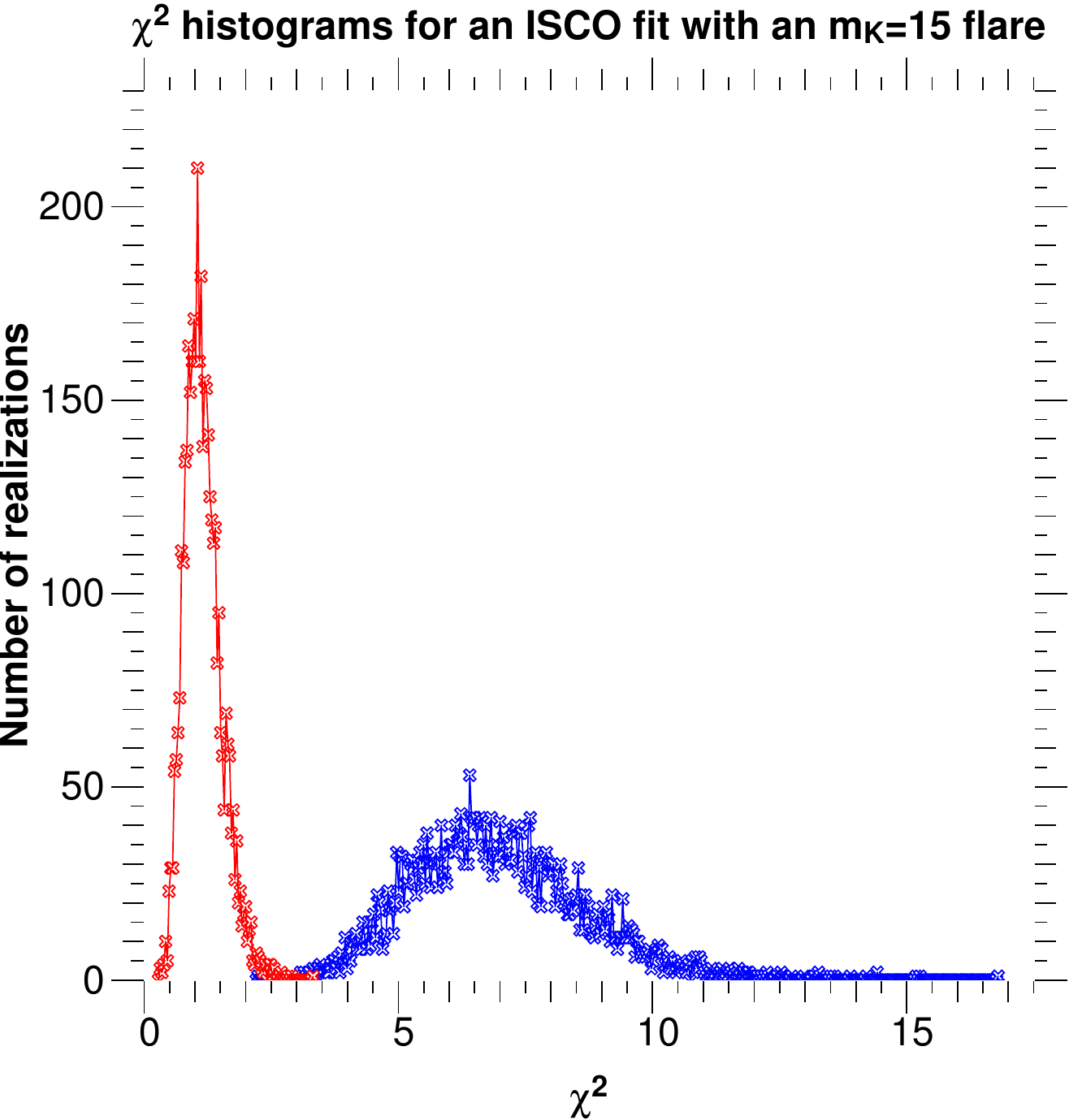}
	\caption{Histograms of reduced $\chi^{2}$ for a non-moving flare (blue) and a flare orbiting around the ISCO (red). This $\chi^{2}$ is the root mean square of the difference between the retrieved positions and a model assuming a circular motion with radius 30 $\mu$as.}
	\label{chi2}
\end{figure*}

Secondly, it can be concluded that when the magnitude of the flare is brighter than 14, it is easy to distinguish between the cases of a moving flare and a non-moving flare. For a flare fainter than $m_{\rm{K}}=15$, it becomes more difficult to conclude because the histograms overlap. However, if we suppose that the flare is indeed due to a hotspot orbiting on the ISCO, it is possible to use the fact that the motion is not random and to fit the positions obtained throughout the night to the last stable orbit. The $\chi^{2}$ given by this procedure is given in Fig.~\ref{chi2} where the fit is performed for an $m_{\rm{K}} = 15$ flare moving on the ISCO, and for a non-moving flare: in the first case, the model is consistent with the data, whereas in the second case, the model is wrong. The question is to determine whether the precision on the flare positions is sufficient to distinguish the two cases. 

It appears that the two distributions are well separated: even for a flare of maximum magnitude $m_{\rm{K}}=15$, it is still possible to determine whether it is moving or not.

However, if a flare of maximum magnitude $m_{\rm{K}}=16$ is considered, the histograms are impossible to distinguish, even if resorting to the ISCO fitting procedure.

It is besides possible to estimate the typical size of the orbit provided the flare is bright enough. Indeed, in the simulation, the flare makes one orbit in 20 min, i.e. it takes 12 snapshots of 100~s each to cover a full orbit. The angle covered between two successive snapshots is of $\theta = \frac{2 \pi}{12}$. Then the theoretical dispersion (the same in both directions) is, with $R_{\rm{isco}}$ being the ISCO radius, and keeping in mind that 15 snapshots are simulated in total:

\begin{eqnarray}
	\label{sigtheo}
\delta_{\rm{theo}}&= & \sqrt{\frac{1}{14}\sum_{k=0}^{14} R_{\rm{isco}}^{2} \rm{cos}^{2}(k\, \theta)} \\ \nonumber
                                & = & 23 \,\mu \rm{as}  \nonumber
\end{eqnarray}
which is quite close to the centre of the red dotted histogram in Fig.~\ref{histodispersion} and~\ref{histodispersionfit} for $m_{\rm{K}}=$13 and 14. Thus, the characteristic size of the orbit is measurable if the flare is as bright at its maximum as $m_{\rm{K}}=14$.

\section{Conclusion}

We have analysed the astrometric information that an instrument such as \textit{GRAVITY} is able to provide when used in `imaging' mode. The accuracy of the positions measured for a field of fixed stars has been investigated. A variable source orbiting on the ISCO of Sgr~A* has then been considered. 

The first study leads to the conclusion that the astrometric precision in pure imaging mode is of the order of, or better than the Schwarzschild radius of the black hole when a single bright enough star is in the field. The limiting brightness of the source is a function of the integration time: $m_{\rm{K}} < 16$ is needed for the whole night of integration to reach this accuracy, whereas $m_{\rm{K}} < 15$ is needed when the integration time is 100 seconds. These results lead to the conclusion that the imaging mode already provides valuable astrometric performance. Besides, when used to bootstrap a fitting procedure, the precision reaches the theoretical limit of the instrument.

The second study makes it clear that it is possible to distinguish between a moving and a non-moving flare on the ISCO, even for flares with a maximum magnitude of only $m_{\rm{K}} = 15$. The positional information can be retrieved in the imaging mode (i.e., without any fit, so without any model). Moreover, the characteristic size of the orbit is at hand provided the flare is as bright as $m_{\rm{K}} = 14$. 

These results are comforting and will be developed in future work. The model of the source is quite crude for the time being and must be refined: for this purpose, a general relativistic ray-tracing code is needed, as well as a more realistic model for the flare. Still, all these results are very encouraging for the GR-related science cases of the instrument: the immediate vicinity of the black hole horizon is within reach. Having access to the motion of the flare would also allow constraining the black hole spin. Indeed, the trajectory of the hotspot is a function of this parameter. However, this constraint will only be possible provided an adequate hotspot model is available.

\section*{Acknowledgments}

This work was supported by grants from R\'egion Ile-de-France.

\appendix

\section{Modelling the interferometric observables}
\label{noise}

\textit{GRAVITY} will use the four VLT UTs to observe Sgr~A*. Each baseline will combine the beams of intensities $I_{p}$ and $I_{q}$ delivered by two telescopes labeled $p$ and $q$. Let the intrinsic visibility modulus and phase of the observed object be $V_{\mathrm{obj}}$ and $\phi_{\mathrm{obj}}$. Let $\delta \phi$ be the phase shift between the two channels from telescope $p$ and $q$. The short-exposure noiseless combined intensity is:

\begin{equation}
	\label{Icomb}
	I(\delta \phi) =  I_{p}+I_{q}+2\sqrt{I_{p} I_{q}}V_{\mathrm{obj}} \mathrm{cos}(\delta \phi + \phi_{\mathrm{obj}} + \phi_{\mathrm{piston,short}})
\end{equation}
where $\phi_{\mathrm{piston,short}}$ is the short-exposure piston phase.

It is sufficient to get four samples of this function in order to retrieve the visibility modulus and the phase of the object. Indeed, if the four following quantities are computed:

\begin{eqnarray}
	\label{ABCD}
A  &=& I(0), \\ \nonumber
B  &=& I(\frac{\pi}{2}), \\ \nonumber
C  &=& I(\pi), \\ \nonumber
D  &=& I(\frac{3\,\pi}{2}), \nonumber
\end{eqnarray}

then it is easy to show that the complex visibility of the object is given by:

\begin{equation}
	\label{visABCD}
	V_{\mathrm{obj}} e^{i\,\phi_{\mathrm{obj}}} = 2\frac{(A-C)-i\,(B-D)}{A+B+C+D}.
\end{equation}

In order to be able to simulate realistic quantities, we now take into account the various sources of noises affecting the different quantities that have been introduced.

\subsection{Detection noise on the four samples A, B, C and D}
\label{abcd}

Let $N_{\rm{ph}}$ be the number of photons arriving from each of the $N_{\rm{tel}}$ telescopes of the interferometer. Each set of $N_{\rm{ph}}$ photons will be dispatched to the $N_{\rm{tel}}-1$ other telescopes. Then $\frac{2N_{\rm{ph}}}{N_{\rm{tel}}-1}$ photons will be present in the baseline between two telescopes. Let  $T$ be the transmission of the instrument multiplied by the quantum yield, estimated to be 0.009. The mean number of photo-electrons $\langle m \rangle$ per sample per baseline is:

\begin{equation}
	\label{photon}
	\langle m \rangle=\frac{2N_{\rm{ph}}T}{4(N_{\rm{tel}}-1)}.
\end{equation}

Hence the shot noise: $\sigma_{\mathrm{shot}}=\sqrt{\langle m \rangle}$. It is assumed that the number of read out noise electrons is equal to $\sigma_{\mathrm{RON}}^{2}=36$. Given the rate of dark current electrons $N_{\mathrm{dark}} = 100\; \rm{s}^{-1}$, the variance of the detection noise per integration time $\tau$ is:

\begin{equation}
	\label{sigmadetec}
	\sigma^{2}_{\mathrm{detec}}=\sigma_{\mathrm{shot}}^{2}+\sigma_{\mathrm{RON}}^{2}+N_{\mathrm{dark}}\,\tau.
\end{equation}

Taking this noise into account, and assuming a gaussian distribution for the various noise contributions, it is possible to compute a realization of the detection noise $n_{A},n_{B},n_{C},n_{D}$ corrupting the four A,B,C,D signals. The signal to noise ratio is lower for fainter sources: it is magnitude dependent.

\subsection{Strehl ratio fluctuation}

It must also be taken into account that the AO-corrected intensities $I_{p}$ and $I_{q}$ are noisy because of Strehl ratio fluctuations. Let $n_{p}$ and $n_{q}$ be realizations of these noises.

\subsection{Visibility modulus and phase noise}
\label{noise_Vphi}

One determination of the visibility modulus and phase using equation~(\ref{visABCD}) is made after having integrated during 100~s. One value of the complex visibility obtained in this way can thus be seen as an average of short exposure complex visibilities which would be obtained after an integration time smaller than the turbulence evolution time (of the order of the ms). Each short exposure complex visibility can be written:

\begin{equation}
	\label{shortV}
	\textbf{V}_{\rm{short}} = V_{\mathrm{obj}} e^{i\,(\phi_{\mathrm{obj}}+\phi_{\mathrm{piston,short}})},
\end{equation}
where $\phi_{\mathrm{piston,short}}$ is the piston phase, which varies at a rate equal to the evolution time of the turbulence.

Averaging these short exposure quantities, the 100~s integrated complex visibility becomes:

\begin{equation}
	\label{longV}
	\textbf{V}_{\rm{sc}} = C\;V_{\mathrm{obj}} e^{i\,(\phi_{\mathrm{obj}}+\phi_{\mathrm{piston}})},
\end{equation}
where $C$ is a contrast parameter between $0$ and $1$, and $\phi_{\mathrm{piston}}$ is the result of the averaging of the different short exposure piston phases. This is the complex visibility obtained at the science beam.

In order to get rid of the piston phase, \textit{GRAVITY} will observe both the science object and a reference star of zero intrinsic phase within the same isopistonic patch. The instrument fringe tracker stabilizes the fringes of the reference and applies the same correction to the object: the science fringe pattern is thus stabilized. However, the two sources being separated on the sky, this procedure is affected by an error depending on the angular separation $\theta$ of the stars: a piston residual thus remains. This anisopistonism error on the location of the source on the sky can be computed for a typical Paranal atmosphere using \citet{delplancke08} and \citet{shao92}:

\begin{equation}
	\label{errfringetracker}
	\sigma_{\mathrm{piston\;res}} [\mathrm{arcsec}] = 370\,\frac{L_{0} [\mathrm{m}]^{\frac{1}{3}}}{B [\mathrm{m}]}\frac{\theta [\mathrm{rad}]}{\sqrt{T [\mathrm{s}]}},
\end{equation}
where $B$ is the baseline, $L_{0}$ is the outer scale of the turbulence and $T$ the integration time.

This allows to compute the residual piston phase due to anisopistonism:

\begin{equation}
	\label{phianiso}
	\phi_{\mathrm{piston\;res}} = \frac{2 \pi}{\lambda} \, B \; \alpha_{\mathrm{piston\;res}},
\end{equation}
where $\alpha_{\mathrm{piston\;res}}$ is a realization of the anisopistonism error on the location of the source on the sky, and $\lambda$ is the observation wavelength.

Moreover, the stabilization of the science fringe pattern on the reference position is done within a certain precision. A residual phase due to the limited performance of the fringe tracking must also be taken into account. It will be denoted $\phi_{\mathrm{tracking\,res}}$.

Finally, the origin of angular positions on sky is by construction the location of the reference star. The phase of the object is thus shifted by $\phi_{\mathrm{met}}=\frac{2\pi}{\lambda}\Delta \mathbf{S}~\mathbf{\cdot}~\mathbf{B}$, where $\mathbf{B}$ is the vector joining the two telescopes (baseline vector) and $\Delta \mathbf{S}$ is the difference between the position vector of the scientific and reference targets on the sky. This shift is determined by the instrument's metrology system. It is affected by an error: $\phi_{\mathrm{met\,res}}~=~0.07 \; \mathrm{rad}$ for an integration time of 30 ms, which is one of the design parameters for \textit{GRAVITY} \citep{rabien08}.

Putting all this together, it becomes possible to express the observed values (denoted with a hat) of the visibility modulus and phase on the science beam:

\begin{eqnarray}
	\label{vphi_sc}
\hat{V}_{\rm{sc}}              &=& C\;V_{\mathrm{obj}} + n_{V_{\rm{sc}}}, \\ \nonumber
\hat{\phi}_{\rm{sc}}  &=& \phi_{\rm{obj}}+\phi_{\mathrm{piston\;res}}+\phi_{\rm{tracking\;res}}+\phi_{\rm{met\; res}}+n_{\phi_{\rm{sc}}}, \nonumber
\end{eqnarray}
where $n_{V_{\rm{sc}}}$ and $n_{\phi_{\rm{sc}}}$ are realizations of the noises affecting the measures of the visibility and the phase.

\subsection{Effective computation in the simulations}

The long-exposure equivalent of equation~(\ref{Icomb}) which takes all noise sources into account is:
	
\begin{eqnarray}
	\label{Icombnoise}
	\hat{I}(\delta \phi) &=&  I_{p}+n_{p}+I_{q}+n_{q} \\ \nonumber
	&+&2\sqrt{(I_{p}+n_{p}) (I_{q}+n_{q})}\,\hat{V}_{\mathrm{sc}}\, \mathrm{cos}(\delta \phi + \hat{\phi}_{\mathrm{sc}}) \\ \nonumber
	&+& n_{\rm{detec}}, \\ \nonumber
\end{eqnarray}
where $n_{\rm{detec}}$ is the realization of the detection noise made explicit in Section~\ref{abcd}.

This quantity can be computed for each of the four samples:

\begin{eqnarray}
	\label{ABCDnoise}
\hat{A}  &=& \hat{I}(0), \\ \nonumber
\hat{B}  &=& \hat{I}(\frac{\pi}{2}), \\ \nonumber
\hat{C}  &=& \hat{I}(\pi), \\ \nonumber
\hat{D}  &=& \hat{I}(\frac{3\,\pi}{2}), \nonumber
\end{eqnarray}

Finally the simulation of the measured complex visibility is computed according to:

\begin{equation}
	\label{visABCDbis}
	\hat{V}_{\mathrm{sc}} e^{i\,\hat{\phi}_{\mathrm{sc}}} = 2\frac{(\hat{A}-\hat{C})-i\,(\hat{B}-\hat{D})}{\hat{A}+\hat{B}+\hat{C}+\hat{D}}.
\end{equation}

In order to work with de-biased quantities, the squared visibility is computed according to:

\begin{equation}
	\label{vis2ABCD}
	\hat{V}_{\mathrm{sc}}^{2} = 4\frac{(\hat{A}-\hat{C})^{2}+(\hat{B}-\hat{D})^{2} - 4 \sigma_{\rm{detec}}^{2}}{(\hat{A}+\hat{B}+\hat{C}+\hat{D})^{2}-4 \sigma_{\rm{detec}}^{2}}
\end{equation}

and the phase is:

\begin{equation}
	\label{phiABCD}
	\hat{\phi}_{\mathrm{sc}} = \rm{atan} \left( \frac{\hat{D}-\hat{B}}{\hat{A}-\hat{C}} \right).
\end{equation}

These values of visibility and phase are delivered by a program that simulates the instrument \textit{GRAVITY}. The noises affecting these quantities are thus at hand, and will be used for the computation of the image from the interferometric observables (see Section~\ref{method}).

\bsp
\label{lastpage}


\begin{thebibliography}{99}
\bibitem[\protect\citeauthoryear{Baganoff et al.}{2001}]{baganoff01} Baganoff F. K. et al., 2001, Nat, 413, 45
\bibitem[\protect\citeauthoryear{Cl\'enet et~al.}{2005}]{clenet05} Cl\'enet Y., Rouan D., Gratadour D., Marco O., L\'ena P., Ageorges N., Gendron E., 2005, A\&A, 439, L9
\bibitem[\protect\citeauthoryear{Delplancke}{2008}]{delplancke08} Delplancke F., 2008, New A Rev., 52, 199
\bibitem[\protect\citeauthoryear{Do et~al.}{2009}]{do09} Do T., Ghez A. M., Morris M. R., Yelda S., Meyer L., Lu J. R., Hornstein S. D., Matthews K., 2009, ApJ, 691, 1021
\bibitem[\protect\citeauthoryear{Dodds-Eden et~al.}{2010}]{doddseden10} Dodds-Eden K. et al., 2010, submitted to ApJ, preprint (astro-ph/1008.1984)
\bibitem[\protect\citeauthoryear{Eckart et~al.}{2009}]{eckart09} Eckart A. et al., 2009, A\&A, 500, 935
\bibitem[\protect\citeauthoryear{Eisenhauer et~al.}{2008}]{eisenhauer08} Eisenhauer F. et al., 2008, Proc. SPIE, 7013E, 69
\bibitem[\protect\citeauthoryear{Genzel et~al.}{2003a}]{genzel03a} Genzel R., Sch\"odel R., Ott T., Eckart A., Alexander T., Lacombe F., Rouan D., Aschenbach B., 2003a, Nat, 425, 934
\bibitem[\protect\citeauthoryear{Genzel et~al.}{2003b}]{genzel03b} Genzel R. et al., 2003b, ApJ, 594, 812
\bibitem[\protect\citeauthoryear{Ghez et~al.}{2004}]{ghez04} Ghez A. M. et al., 2004, ApJ, 601, L159
\bibitem[\protect\citeauthoryear{Ghez et~al.}{2008}]{ghez08} Ghez A. M. et al., 2008, ApJ, 689, 1044
\bibitem[\protect\citeauthoryear{Gillessen et~al.}{2009}]{gillessen09} Gillessen S., Eisenhauer F., Trippe S., Alexander T., Genzel R., Martins F., Ott T., 2009, ApJ, 692, 1075
\bibitem[\protect\citeauthoryear{Gillessen et~al.}{2010}]{gillessen10} Gillessen S. et al., 2010, Proc. SPIE, Vol. 7734, 77340Y
\bibitem[\protect\citeauthoryear{Hamaus et~al.}{2009}]{hamaus09} Hamaus N., Paumard T., M\"uller T., Gillessen S., Eisenhauer F., Trippe S., Genzel R., 2009, ApJ, 692, 902
\bibitem[\protect\citeauthoryear{Haubois et~al.}{2009}]{haubois09} Haubois X. et al., 2009, A\&A, 508, 923
\bibitem[\protect\citeauthoryear{H\"ogbom et~al.}{1974}]{hogbom74} H\"ogbom J. A., 1974, A\&AS, 15, 417
\bibitem[\protect\citeauthoryear{Houairi et~al.}{2008}]{houairi08} Houairi K., Cassaing F., Perrin G., Eisenhauer F., Brandner W., Straubmeier C., Gillessen S., 2008, Proc. SPIE, Vol. 7013, 70131B
\bibitem[\protect\citeauthoryear{Le~Besnerais et~al.}{2008}]{lebesnerais08} Le Besnerais G., Lacour S., Mugnier L. M., Thi\'ebaut E., Perrin G., Meimon S., 2008, IEEE Journal of Selected Topics in Signal Processing, 2, 767
\bibitem[\protect\citeauthoryear{Markoff et~al.}{2001}]{markoff01} Markoff S., Falcke H., Yuan F., Biermann P. L., 2001, A\&A, 379, L13
\bibitem[\protect\citeauthoryear{Meimon, Mugnier~\&~Le~Besnerais}{Meimon et~al.}{2009}]{meimon09} Meimon S., Mugnier L. M., Le Besnerais G., 2009, J. Opt. Soc. Am. A, 26, 108
\bibitem[\protect\citeauthoryear{Paumard et~al.}{2006}]{paumard06} Paumard T. et al., 2006, ApJ, 643, 1011
\bibitem[\protect\citeauthoryear{Paumard et~al.}{2008}]{paumard08} Paumard T. et al., 2008, Proc. ESO Workshop, Vol. 2008, 313
\bibitem[\protect\citeauthoryear{Pott et~al.}{2008}]{pott08} Pott J.-U. et al., 2008, preprint (astro-ph/0811.2264)
\bibitem[\protect\citeauthoryear{Rabien et~al.}{2008}]{rabien08} Rabien S. et al., 2008, Proc. SPIE, Vol. 7013, 70130I
\bibitem[\protect\citeauthoryear{Rubilar~\&~Eckart}{2001}]{rubilar01} Rubilar G. F., Eckart A., 2001, A\&A, 374, 95
\bibitem[\protect\citeauthoryear{Shannon}{1949}]{shannon49} Shannon C. E., 1949, Proc. Institute of Radio Engineers, Vol. 37, 10
\bibitem[\protect\citeauthoryear{Shao~\&~Colavita}{1992}]{shao92} Shao M., Colavita M. M., 1992, A\&A, 262, 353
\bibitem[\protect\citeauthoryear{Sch\"odel et~al.}{2002}]{schodel02} Sch\"odel R. et al., 2002, Nat, 419, 694
\bibitem[\protect\citeauthoryear{Tagger~\&~Melia}{2006}]{tagger06} Tagger M., Melia F., 2006, ApJ, 636, L33
\bibitem[\protect\citeauthoryear{Thi\'ebaut}{2008}]{thiebaut08} Thi\'ebaut E., 2008, Proc. SPIE, Vol. 7013, 70131I-12
\bibitem[\protect\citeauthoryear{Will}{2008}]{will08} Will C. M., 2008, ApJ, 674, L25
\bibitem[\protect\citeauthoryear{Yuan, Quataert~\&~Narayan}{Yuan et~al.}{2004}]{yuan04} Yuan F., Quataert E., Narayan R., 2004, ApJ, 606, 894
\bibitem[\protect\citeauthoryear{Yusef-Zadeh et~al.}{2006}]{yusefz06} Yusef-Zadeh F., Roberts D., Wardle M., Heinke C. O., Bower G. C., 2006, ApJ, 650, 189
\end{thebibliography}
\end{document}